\pdfoutput=1
\documentclass[aps, nofootinbib,subeqnt]{revtex4}
\usepackage{graphicx}% Include figure files
\usepackage{amsmath}
\usepackage{amsfonts}
\usepackage{amssymb}
\usepackage{appendix}
\usepackage{mathtools}
\usepackage{comment}
\usepackage{color}
\usepackage{slashed}
\usepackage{subfigure}

\begin{document}

%\title{$M_{T2}$ and Light Invisible Particles}
\title{The Lightest Massive Invisible Particles at the LHC}
\date{\today}

\author{Andr\'{e} de Gouv\^{e}a} \affiliation{Northwestern University, Department of Physics \& Astronomy, 2145 Sheridan Road, Evanston, IL 60208, USA}
\author{Andrew C.~Kobach} \affiliation{Northwestern University, Department of Physics \& Astronomy, 2145 Sheridan Road, Evanston, IL 60208, USA}
\date{\today}

\begin{abstract}
The observation of new physics events with large missing transverse energy at the LHC would potentially serve as evidence for the direct production of dark matter. A crucial step toward verifying such evidence is the measurement of the would-be dark matter mass. If, for example, the invisible particles are found to have masses consistent with zero, it may prove very challenging to ascertain whether light dark matter or neutrinos are being observed.  We assume that new invisible particles are pair-produced in a $t\bar{t}$-like topology and use two $M_{T2}$-based methods to measure the masses of the particles associated with the missing energy.  Instead of simulating events and backgrounds, we estimate the uncertainty associated with measuring the mass of the invisible particle by assuming a fixed value of the uncertainty associated with the location of the $M_{T2}$ endpoint.  We find that if this uncertainty associated with measuring the $M_{T2}$ endpoints is, quite optimistically, $\mathcal{O}(1 \text{ GeV})$, the invisible particles must have masses greater than $\mathcal{O}(10 \text{ GeV})$ so they can be distinguished from massless ones at 95\% CL.    If the results from the CoGeNT, DAMA/LIBRA, and CRESST experiments have indeed revealed the existence of light dark matter with mass $\mathcal{O}(\text{10 GeV})$, our results suggest that it may be difficult for the LHC to distinguish dark matter from neutrinos solely via mass measurements.
\end{abstract}
 

\maketitle

% INTRODUCTION
\section{Introduction}
One of the main goals of high-energy accelerator experiments like the LHC is the discovery, via direct production and subsequent detection of the associated decay products, of degrees of freedom beyond those present in the standard model. An advantage of direct production is that the various properties of the new particles, including their masses, can be measured, often with very good precision. Different methods exist for measuring the masses of the particles involved in different production and decay topologies~\cite{Gripaios:2011na, Barr:2010zj}, but a unique problem arises when one or more of the final-state particles are ``invisible," i.e., they do not interact directly with the detector.  At a hadron collider, the existence of invisible particles can be inferred only via momentum conservation in the plane transverse to the beamline; only the vector sum of the  transverse momenta associated with invisible particles can be reconstructed.  If two missing particles in a single event each originate from identical decay chains, e.g., a $t\bar{t}$-like topology\footnote{We define a $t\bar{t}$-like event as any decay topology similar to the standard model (SM) leptonic decay of two top quarks: $t\bar{t}\rightarrow W^+ b W^-\bar{b}\rightarrow \ell^+ \nu_\ell b \ell^- \bar{\nu}_\ell \bar{b}$.  We call the top quark the parent particle, the $W$ boson the intermediate particle, and the neutrino is the invisible particle.  All of these particles are considered to be potentially massive, while all other particles in the decay chain are assumed to be massless.}, the $M_{T2}$ variable~\cite{Lester:1999tx} is a useful tool for extracting the masses of the parent, intermediate, and invisible particles in the decay chain.  Remarkably, this includes the mass of the particles associated with the missing energy in the event~\cite{Barr:2007hy, Cho:2007qv, Cho:2007dh, Burns:2008va}.   A $t\bar{t}$-like topology is one of the best-suited topologies for measuring the mass of the invisible particles. 

Two popular $M_{T2}$-based methods for extracting the masses of the particles in a $t\bar{t}$-like decay topology are reconstructing the $M_{T2}$ kink~\cite{Barr:2007hy, Cho:2007qv, Cho:2007dh} and the $M_{T2}$ subsystems method~\cite{Burns:2008va}.  The $M_{T2}$ kink method involves measuring $M_{T2}$ endpoints, which we call $M_{T2}^\text{max}$, for different values of an ansatz for the invisible particles' mass~\cite{Barr:2007hy, Cho:2007qv, Cho:2007dh}.  Analytical expressions can be fit to the distribution of $M_{T2}^\text{max}$ as a function of the input ansatz, and the masses of the particles in the decay chain can be determined simultaneously from this fit.  A kink exists in the $M_{T2}^\text{max}$ distribution at the mass of the invisible particles, but as the mass of the invisible particles becomes light, it may be difficult to determine that the location of the kink is non-zero, given experimental uncertainties.  Another $M_{T2}$-based method involves studying subsystems of the decay chain and measuring three kinematic endpoints~\cite{Burns:2008va}.  With the analytical expressions for the endpoints of each subsystem, one can simultaneously solve for the masses of the particles in the decay chain.  If the mass of the invisible particle is light, the uncertainty for solving for its mass can become large.  Because of the experimental uncertainties associated with these $M_{T2}$ methods, there is a minimum mass of the invisible particle above which it can distinguished from a massless particle at 95\% C.L.  

If a large missing-energy signal is discovered at the LHC, it is possible that this signal could be due not to dark matter, but to anomalous production of neutrinos~\cite{Chang:2009dh}.  Measuring the mass of the invisible particles is the only model-independent way to distinguish massive invisible particles, e.g. dark matter, from those that are effective massless, e.g., neutrinos.  We employ the $M_{T2}$ kink and the $M_{T2}$ subsystems methods to determine how heavy the invisible particles must be in order to be distinguishable from a massless hypothesis, which, in essence, is an estimation of the lower bound on detectable dark matter at a hadron collider.  

Much of the literature to date uses SUSY models when addressing mass determination with $M_{T2}$ methods, taking the LSP to be $\mathcal{O}(100\text{ GeV}$)~\cite{Cho:2007qv, Cho:2007dh, Nojiri:2008hy, Hamaguchi:2008hy, Alwall:2009zu, Konar:2009qr, Cho:2009wh,  Agashe:2010tu, Ajaib:2010ne, Chen:2010ek, Choi:2010dk, Cohen:2010wv, Park:2011uz}.  The motivations for considering dark matter to have these masses in the context of the MSSM and in light of experimental constraints are described in Ref.~\cite{Belanger:2012jn} and the references found therein.   Data from the CoGeNT~\cite{Aalseth:2010vx}, DAMA/LIBRA~\cite{Bernabei:2008yi,Bernabei:2010mq}, and CRESST~\cite{Angloher:2011uu} experiments, on the other hand, hint at the existence of light dark matter (mass of order a few to 10 GeV).  Only a few examples exist in the literature which study light dark matter properties, including masses, at hadron colliders, in particular, Ref.~\cite{Belanger:2011ny}.   In that analysis, the authors consider a light sneutrino with mass $\mathcal{O}(10\text{ GeV})$ and demonstrate that when using $M_{T2}$-based methods, the uncertainty associated with the measured mass of the sneutrino can be relatively large.  They consider various experimental effects, appropriate for their analysis, e.g., backgrounds, combinatorial ambiguity, initial-state radiation, etc.  Since all of these effects make it more difficult to locate the $M_{T2}$ endpoint, here we simply quote our results as a function of uncertainty with measuring it.  We also take care to estimate the correlations between $M_{T2}$ endpoints as a function of the input ansatz mass, an effect not considered in Ref.~\cite{Belanger:2011ny}.  We believe that by ignoring these correlations one may significantly underestimate the uncertainties associated with measuring the mass of the invisible particles.

The outline of our analysis is as follows:~in Section~\ref{mt2variable}, we define and review the $M_{T2}$ variable and express the previously-derived~\cite{Barr:2007hy, Cho:2007qv, Cho:2007dh} expressions for its endpoints.  In Sections~\ref{kinkmethod} and~\ref{subsmethod}, we create pseudo-data and use the $M_{T2}$ kink and subsystems methods, respectively, to determine the mass that the invisible particle must have in order to distinguish it from a massless particle at 95\% C.L.\footnote{We use the term ``C.L." to describe the uncertainty associated with the sampling method.}  We do this for various masses of the parent and intermediate particles, and different uncertainties associated with the determination of the $M_{T2}$ endpoints.  We conclude and discuss our results in Section~\ref{conclusion}.

% MT2 VARIABLE
\section{The $M_{T2}$ Variable} 
\label{mt2variable}
Consider a general decay chain where a single massive particle, $A$, with mass $m_A$, decays to $N$ massless, visible final-state particles and one potentially massive, invisible particle with physical mass $m$.  The individual three-momenta of the visible particles, $\vec{p}_i$, are measured, and their four-momenta, $p^\mu_i$, are inferred with the massless approximation.\footnote{The results presented here also apply for massive visible particles, as long as these are properly identified.}  The invariant mass of the visible system, $M_\text{vis}$, is defined as 
\begin{eqnarray}{}
M^2_\text{vis} \equiv P_\mu P^\mu, &&P^\mu \equiv \displaystyle\sum_{i}^{N} p_i^\mu.
\end{eqnarray}
At a hadron collider, the sum of the transverse momentum of all final-state particles in the event is zero (to a good approximation), and the transverse momentum of the invisible particle, $\vec{\slashed{p}}_T$, can be inferred.  We adopt an ansatz for the mass of the invisible particle, $\tilde{m}$, and define the transverse mass variable $M_T$, the square of which is defined as

\begin{equation}
\label{mtdef}
M_T^2(P^\mu, \vec{\slashed{p}}_T; \tilde{m}) \equiv  M_\text{vis}^2 + \tilde{m}^2 + 2\left(\sqrt{M_\text{vis}^2 + |\vec{P}_T|^2} \sqrt{\tilde{m}^2 + |\vec{\slashed{p}}_T|^2} - \vec{P}_T \cdot \vec{\slashed{p}}_T \right).
\end{equation}
A distribution of $M_T$ values displays a ``kinematic endpoint" or ``edge" at the value of $m_A$ when $\tilde{m} = m$.  

Consider now an event where a pair of $A$'s is created.  Each parent particle and their daughters belong to a ``branch" or ``decay chain", where we add the label ``(1)" and ``(2)" to distinguish between the respective decay chains.  The two parent particles, $A^{(1)}$ and $A^{(2)}$, eventually decay to $N^{(1)}$ and $N^{(2)}$ visible effectively-massless particles and a potentially-massive invisible particle with mass $m$.\footnote{It is unnecessary to provide any further details concerning the topology of the decay chain at this point.}  The measured missing transverse momentum, $\vec{\slashed{p}}_T$, is the vector sum of the transverse momenta of the two invisible particles.  Because there are two invisible particles in the final state, the $M_T$ variable does not provide information concerning the masses of the particles in the decay chain.  However, a generalized $M_T$ variable, called $M_{T2}$, can be introduced~\cite{Lester:1999tx}, and because we do not know how the transverse momentum is shared between the invisible particles, the square of $M_{T2}$ is defined as 

\begin{equation}
\label{mt2def}
M_{T2}^2(\tilde{m}) \equiv \min_{\vec{k}^{(1)}_T+\vec{k}^{(2)}_T = \vec{\slashed{p}}_T} \max \left[ M_T^2\left(P^{\mu (1)}, \vec{k}^{(1)}_T; \tilde{m}\right), M_T^2\left(P^{\mu (2)}, \vec{k}^{(2)}_T; \tilde{m}\right) \right],
\end{equation}
where $\vec{k}^{(1)}_T$ and $\vec{k}^{(2)}_T$ are free parameters over which the function is minimized, subject to the constraint $\vec{k}^{(1)}_T+\vec{k}^{(2)}_T = \vec{\slashed{p}}_T$.  The distribution of $M_{T2}$ exhibits an endpoint, $M_{T2}^\text{max}$, at $m_A$ when $\tilde{m} = m$.  If $\tilde{m}\neq m$, the $M_{T2}$ endpoint still exists, but its location does not equal $m_A$.   In general, the change in the value of $M_{T2}^\text{max}$ as a function of $\tilde{m}$ contains the information of the masses of the particles in the decay chains.  A remarkable quality of the $M_{T2}$ variable is, for some topologies, the analytical form that describes $M_{T2}^\text{max}$ when $\tilde{m} > m$ is different from that when $\tilde{m} < m$.  While these functions are continuous for all $\tilde{m}$, their first derivatives are not, and a ``kink" exists in the distribution of $M_{T2}^\text{max}(\tilde{m})$ at $\tilde{m} = m$ \cite{Cho:2007qv, Cho:2007dh}.  By analyzing the shape of the $M_{T2}^\text{max}$ distribution as a function of $\tilde{m}$, one can determine, in principle, the masses of the particles in the decay chains. 

\begin{figure}[htbp]
\begin{center}
\includegraphics[width=0.45\textwidth]{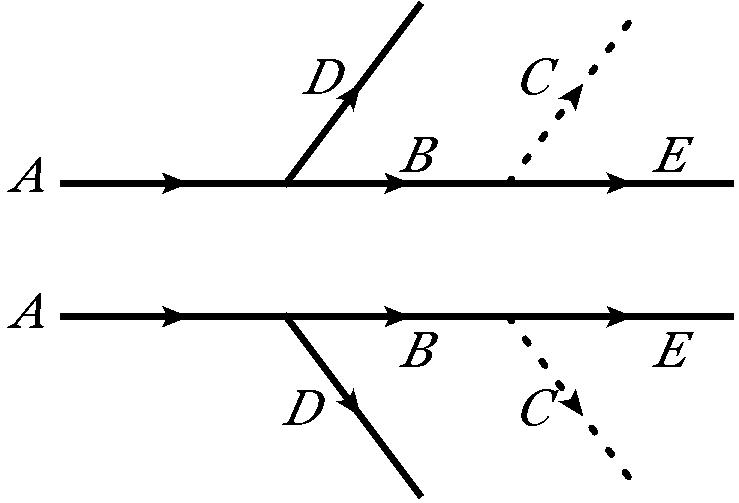}
\caption{The decay topology of the pair production of parent particle, $A$, which decays into two visible effectively-massless final-state particles, $D$ and $E$, and an invisible potentially-massive particle, $C$.  Here, $B$ is considered to be on-shell.}
\label{ttbardecay}
\end{center}
\end{figure}

For the rest of our discussions, we will assume a $t\bar{t}$-like topology, as shown in Fig.~\ref{ttbardecay}: $A^{(i)}\rightarrow B^{(i)} D^{(i)}$, followed by $B^{(i)}\rightarrow C^{(i)} E^{(i)}$, where $i=1,2$ and $C$ is considered to be the potentially-massive invisible particle.  We make no further assumptions regarding what particle types are $D$ or $E$, only that they are effectively-massless and their three-momenta can be reconstructed.  Particles $A$, $B$, and $C$ have mass $m_A$, $m_B$ and $m_C$, respectively, and we assign the ansatz $\tilde{m}_C$ for the mass of the invisible particles.  
The analytical expressions for $M_{T2}^\text{max}(\tilde{m}_C)$ for the $t\bar{t}$-like topology are, in the limit of no initial-state radiation (ISR)~\cite{Cho:2007dh, Barr:2007hy},
\begin{eqnarray}
\label{mt2max}
M_{T2}^\text{max}(\tilde{m}_C) &\equiv& \left\{ \begin{array}{l l} \frac{m_A^2-m_C^2}{2m_A} + \sqrt{\left(\frac{m_A^2-m_C^2}{2m_A}\right)^2 + \tilde{m}_C^2}, & \tilde{m}_C \leq m_C, \\ \frac{m_A^2-m_B^2}{2m_A} + \frac{m_A}{2}\left(1- \frac{m_C^2}{m_B^2}\right) + \sqrt{\left[ \frac{m_A^2-m_B^2}{2m_A} - \frac{m_A}{2}\left(1- \frac{m_C^2}{m_B^2}\right) \right]^2 + \tilde{m}_C^2}, & \tilde{m}_C \geq m_C. \end{array}\right. 
\end{eqnarray}
By fitting the distribution of $M_{T2}^\text{max}$ as a function of $\tilde{m}_C$ with the expression for $M_{T2}^\text{max}(\tilde{m}_C)$ in Eq.~(\ref{mt2max}), one can simultaneously solve for $m_A$, $m_B$, and $m_C$.  

In the absence of experimental effects, the $M_{T2}$ endpoint is a sharp and easily distinguishable feature.  However, given a realistic collider environment, finding the kinematic endpoint of the $M_{T2}$ distribution can be difficult~\cite{Curtin:2011ng}.  The endpoint feature can be obfuscated by decay widths, finite detector resolutions, and lack of statistics around the kinematic endpoint.  Additionally, it is often experimentally difficult to distinguish different types of final-state jets, e.g., if a reconstructed jet is due to a gluon or quark, so if there are jets in the final state of the pair-produced decay chain, the $M_{T2}$ endpoints can be contaminated with initial-state radiation jets~\cite{Krohn:2011zp, Nojiri:2010mk}.  There can also exist combinatorial ambiguities associated with which jet is to be paired with which decay chain~\cite{Baringer:2011nh, Choi:2011ys, Rajaraman:2010hy}.  For the purposes of our study, we encapsulate these experimental effects by quoting our results as a function of the uncertainty associated with determining the location of an $M_{T2}$ endpoint, $\sigma_E$.  In particular, we assume that $\sigma_E$ will be the same for all values of $\tilde{m}_C$, $m_A$, and $m_B$.  While it is a simplifying assumption to distill these experimental effects to a single number, our goal is to estimate the precision with which a collider experiment can measure the mass of a final-state invisible particle, instead of performing a detailed study of specific experimental effects.  We will present our results for $\sigma_E=1$ and 5 GeV (assuming it is a convolution of statistical and systematic uncertainties), which we consider to be optimistic estimates for the uncertainties associated with the capabilities of the LHC~\cite{Chatrchyan:2013boa}.  In particular, the analysis in Ref.~\cite{Chatrchyan:2013boa}, which uses a clean sample of SM $t\bar{t}$ events, measures the location of the three $M_{T2}$ subsystems endpoints with a precision of $\sigma_{E_i} = \mathcal{O}(\text{10 GeV})$~\cite{eggert}.

% KINK METHOD
\section{The $M_{T2}$ Kink Method and Light Dark Matter}
\label{kinkmethod}
If the mass of the invisible particle is small, i.e., small relative to the $M_{T2}$ endpoint uncertainty, $\sigma_E$,  it may be difficult to distinguish a massive invisible particle from one that is massless.  To estimate how heavy the dark matter must be in order to be distinguished from anomalous neutrino production, we must study the uncertainties associated with measuring the mass of the invisible particle using the $M_{T2}$ kink method. 
To do this, one could consider producing Monte Carlo (MC) events of a $t\bar{t}$-like topology with a detector simulation and using a procedure to measure the $M_{T2}^\text{max}$ distribution as a function of $\tilde{m}_C$.  However, because of the significant amount of time it would take to create and analyze such MC samples, we choose to simulate simplified pseudo-data of the $M_{T2}^\text{max}(\tilde{m}_C)$ distribution.  A full description of how we created simplified pseudo-data can be found in Appendix~\ref{apa}.  We take note that the individual values of $M_{T2}^\text{max}(\tilde{m}_C)$ are highly correlated between values of $\tilde{m}_C$ that are close together.  This is easily understood since the events that populate the $M_{T2}$ kinematic endpoint for a given $\tilde{m}_C$ are mostly the same, regardless of the choice of $\tilde{m}_C$.\footnote{The endpoint of the $M_{T2}$ distribution is due to a particular momentum configuration of the final-state particles in the decay chains~\cite{Cho:2007qv, Cho:2007dh}.}  For simplicity, we assume that this correlation does not depend on the physical masses of the decay topology,\footnote{In general, highly-correlated values of $M_{T2}^\text{max}$ for different values of $\tilde{m}_C$ involve a high percentage of the $M_{T2}^\text{max}$ region from ``unbalanced solutions" of $M_{T2}$ (the definition of which can be found in Refs.~\cite{Cho:2007dh, Barr:2003rg}).} and we choose to sample $M_{T2}^\text{max}(\tilde{m}_C)$ in 0.25 GeV steps of $\tilde{m}_C$.  We find that a smaller step size of $\tilde{m}_C$ yields endpoints that are too correlated for adjacent values of $\tilde{m}_C$, and a larger step size would hinder the ability to resolve the position of the kink when it has a value close to zero.  

When simulating the simplified pseudo-data, we take care to estimate the positive correlations between $M_{T2}$ endpoint measurements. Ignoring to do so would imply that the uncertainty associated with the location of the $M_{T2}$ kink depends on the choice of step size of $\tilde{m}_C$, which is arbitrary, and the values of $m_C^\text{min}$ can be significantly underestimated.   For example, if we repeat our analysis ignoring the correlations between the $M_{T2}$ endpoints for values of $\tilde{m}_C$ that are 0.25 GeV apart, then the values for $m_C^\text{min}$ is underestimated by almost a factor of two.  We note that the presence of these correlations is independent from the treatment of the uncertainties associated with measuring the location of the $M_{T2}$ endpoint.   

The full summary of our estimation of the correlation between $M_{T2}(\tilde{m}_C)$ endpoints can be found in Appendix~\ref{apa}.  We generate fifty thousand pseudo-data $M_{T2}^\text{max}(\tilde{m}_C)$ distributions for different values of $m_C$, fitting them with the analytical functions for $M_{T2}^\text{max}$ in Eq.~(\ref{mt2max}), for which the physical masses of the decay topology are the fitting parameters.  For each fitted distribution of $M_{T2}^\text{max}$, we histogram the fifty thousand best fit values of $m_C$ (which explicitly marginalizes over the uncertainties associated with the $m_A$ and $m_B$ fitting parameters), and from the width of this histogram, we estimate the 95\% C.L. associated with the value of $m_C$.

\begin{figure}[htbp]
\begin{center}
\subfigure[]{\label{sig11}\includegraphics[width=0.45\textwidth]{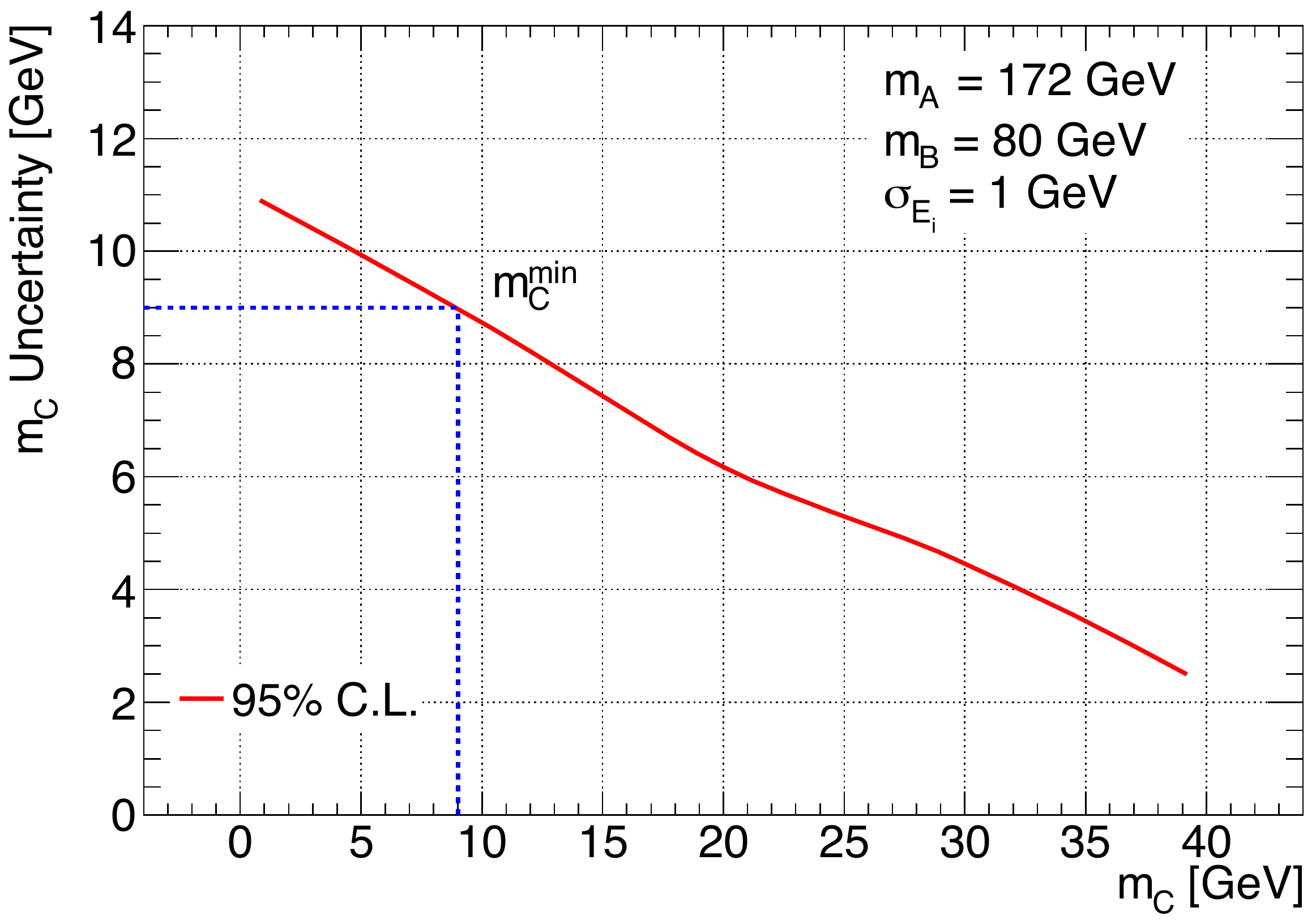}}
\subfigure[]{\label{sig51}\includegraphics[width=0.45\textwidth]{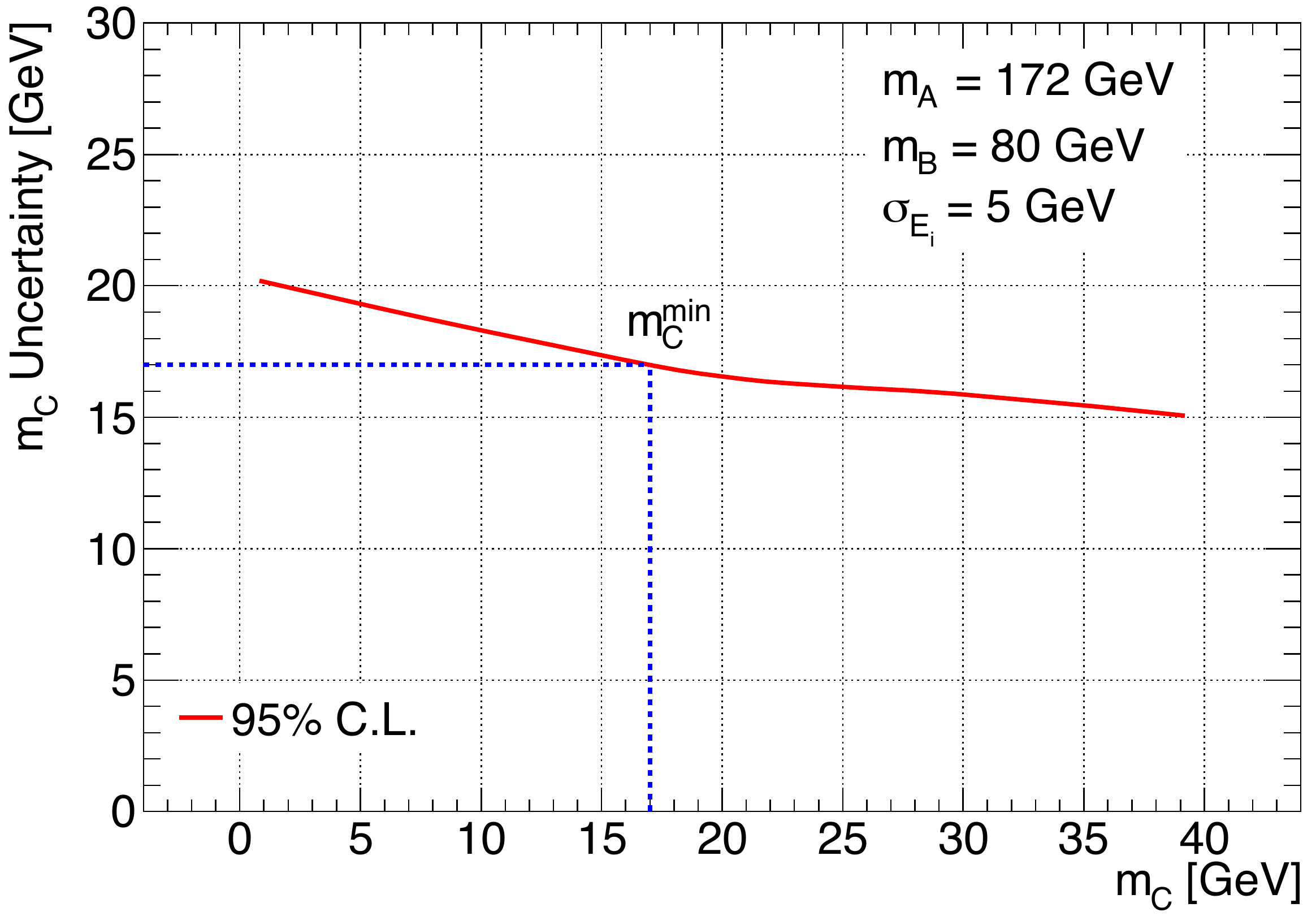}}
\caption{The 95\% C.L. for the mass of $m_C$ for $m_A=172$ GeV and $m_B=80.4$ GeV with (a) $\sigma_{E_i}=1$ GeV and (b) $\sigma_{E_i}=5$ GeV, using the $M_{T2}$ kink method.  The variable $m_C^\text{min}$ is the value of mass of $m_C$ at which it can be distinguished from zero at 95\% C.L.}
\label{kink1}
\end{center}
\end{figure}

\begin{figure}[htbp]
\begin{center}
\subfigure[]{\label{sig12}\includegraphics[width=0.45\textwidth]{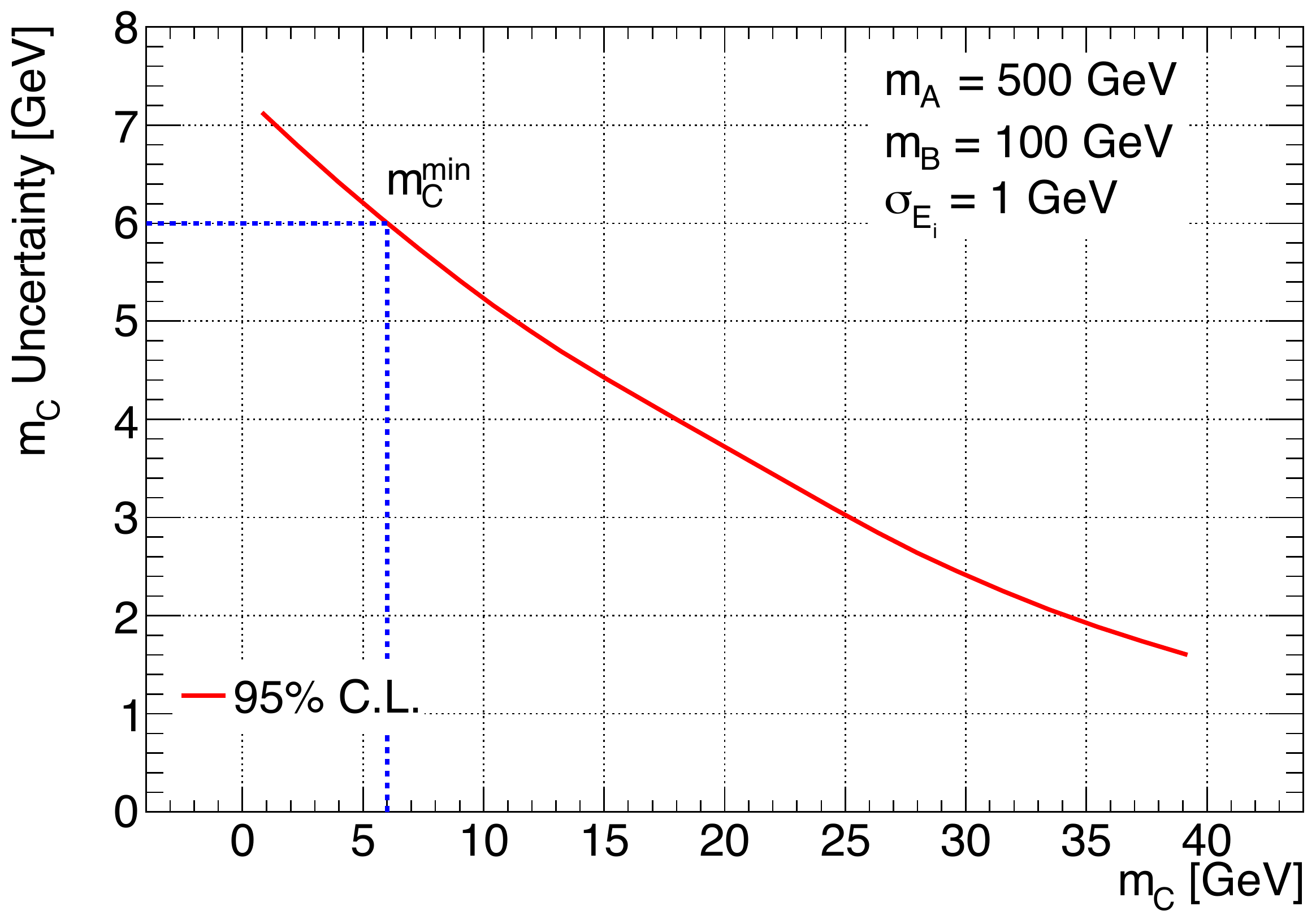}}
\subfigure[]{\label{sig52}\includegraphics[width=0.45\textwidth]{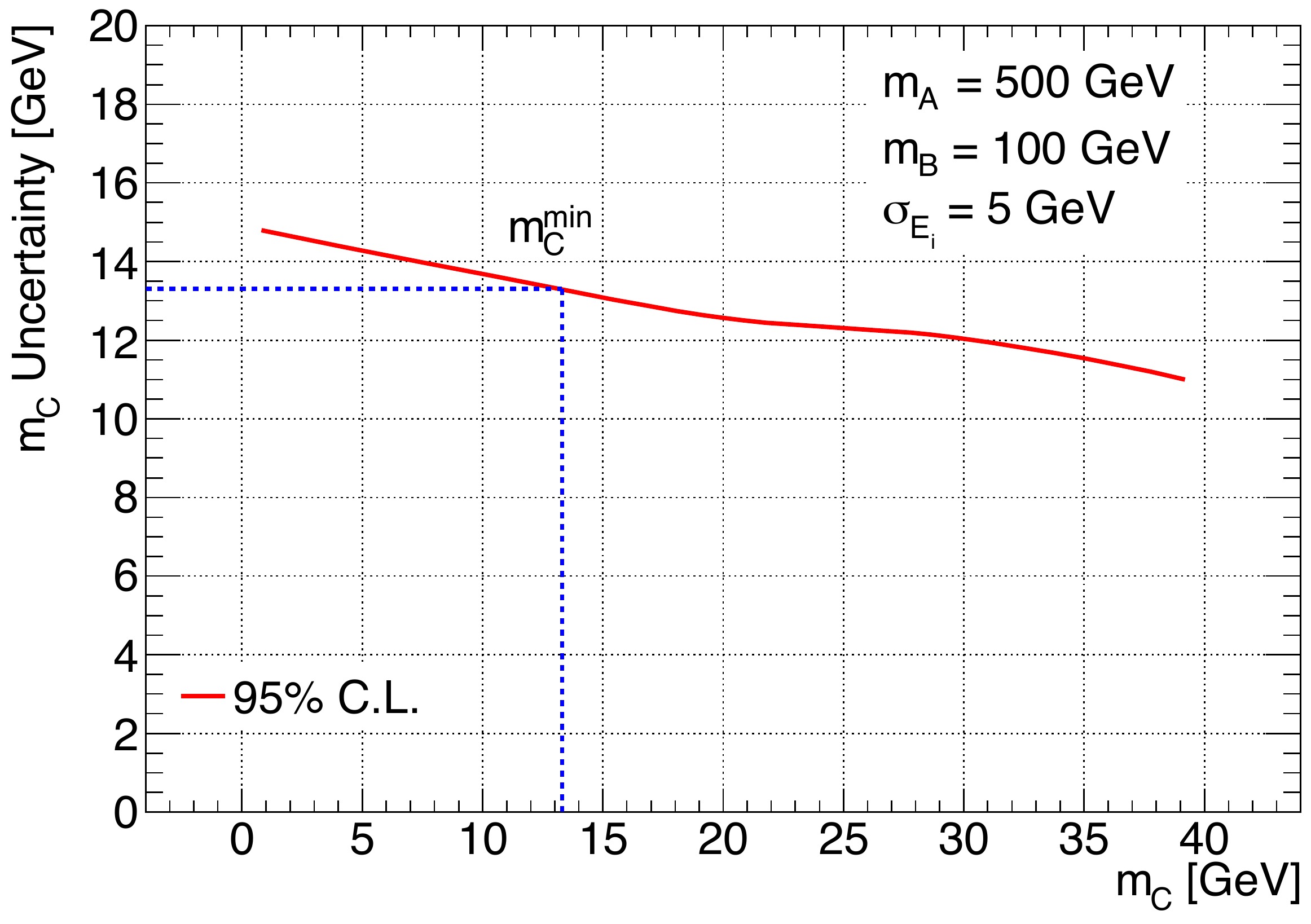}}
\caption{As in Fig.~\ref{kink1}, for $m_A=500$ GeV and $m_B=100$ GeV.}
\label{kink2}
\end{center}
\end{figure}

\begin{figure}[htbp]
\begin{center}
\subfigure[]{\label{sig13}\includegraphics[width=0.45\textwidth]{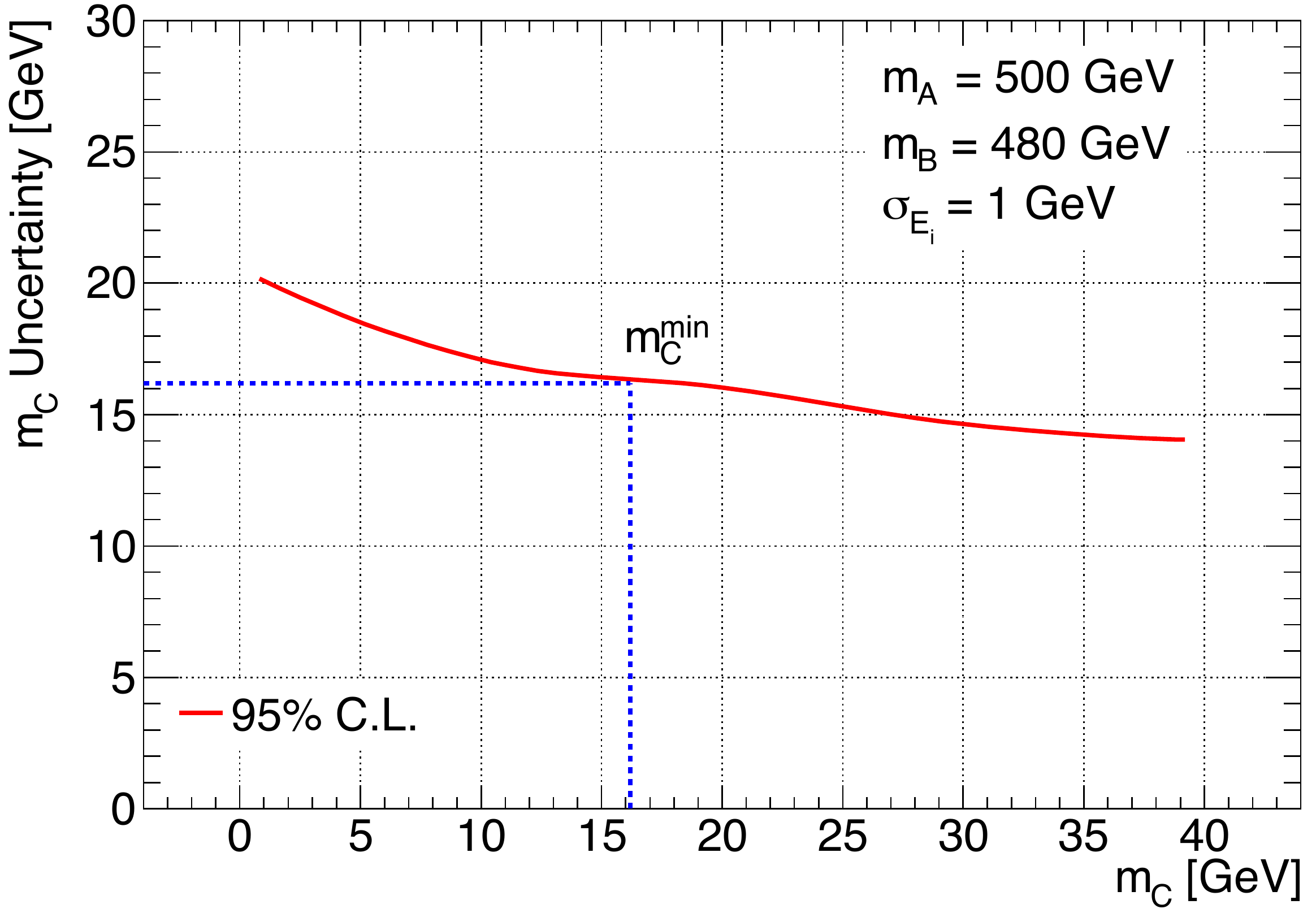}}
\subfigure[]{\label{sig53}\includegraphics[width=0.45\textwidth]{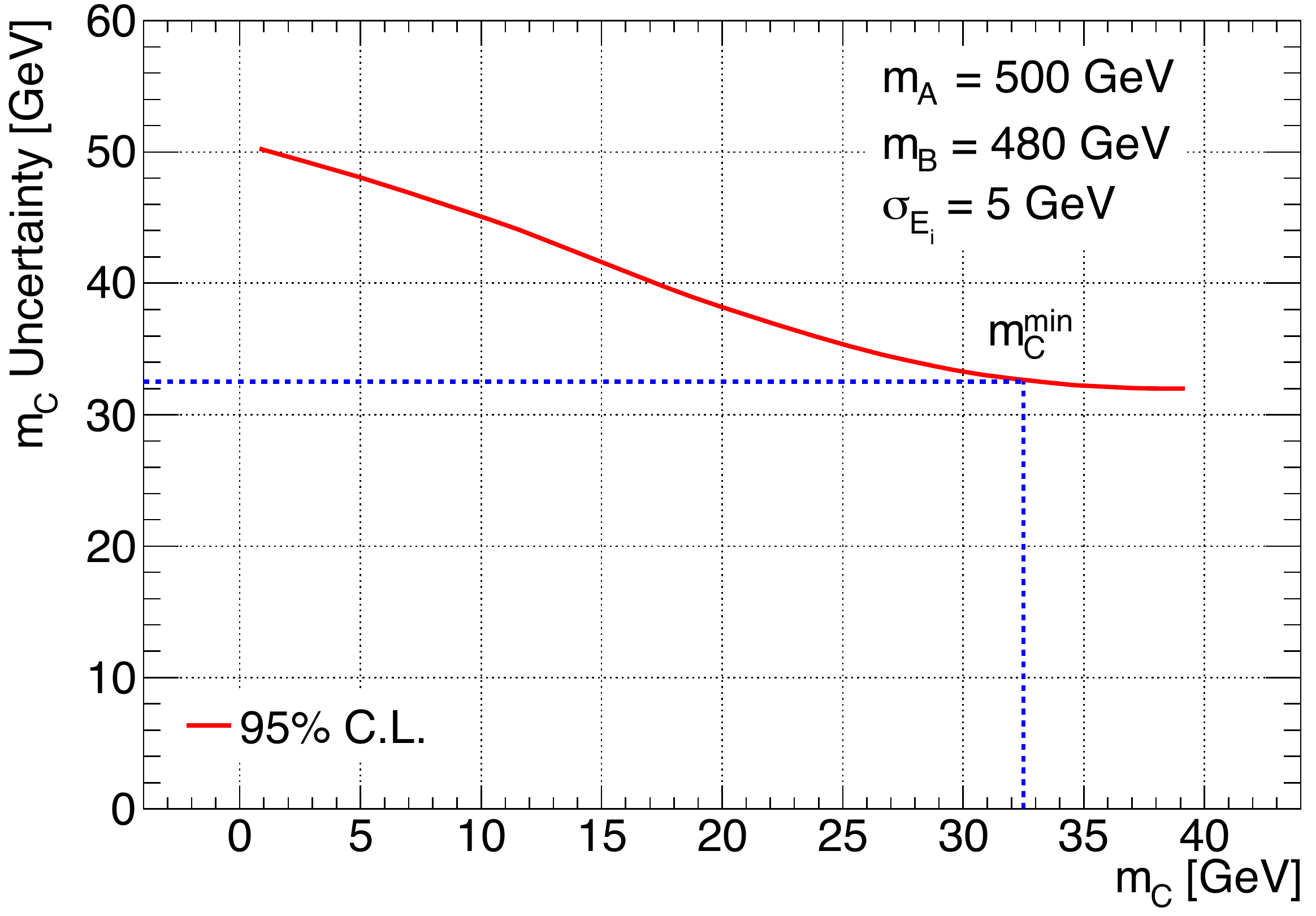}}
\caption{As in Fig.~\ref{kink1}, for $m_A=500$ GeV and $m_B=480$ GeV.}
\label{kink3}
\end{center}
\end{figure}

To draw similarities from a SM production of $t\bar{t}$ events, i.e., for $m_A = 172$ GeV, $m_B = 80.4$ GeV (while still letting $m_C$ float), the uncertainties of the measurement of $m_C$ given $\sigma_{E_i} = 1$ GeV and $\sigma_{E_i} = 5$ GeV are shown in Fig.~\ref{kink1}.  From these figures, one can determine what the mass of $m_C$ must be in order to distinguish it from $m_C=0$ at 95\% C.L.  If we call this value $m_C^\text{min}$, then $m_C^\text{min} \approx 9$ GeV if $\sigma_E = 1$ GeV, and $m_C^\text{min} \approx 17$ GeV if $\sigma_E = 5$ GeV.  We also estimate $m_C^\text{min}$ when $m_A = 500$ GeV and $m_B=100$ GeV, as shown in Fig.~\ref{kink2}, and when $m_A = 500$ GeV and $m_B=480$ GeV, as shown in Fig.~\ref{kink3}.   
These results for the value of $m_{C}^\text{min}$, given the values of $m_A$, $m_B$, and $\sigma_E$, can be understood as a lower bound on a measured non-zero mass of dark matter.

%

% SUBSYSTEMS METHOD
\section{The $M_{T2}$ Subsystems Method and Light Dark Matter}
\label{subsmethod}
In order to address some of the combinatorial difficulties associated with the $M_{T2}$ kink method, the authors of Ref.~\cite{Burns:2008va} developed what is known as the $M_{T2}$ subsystems method.  To simultaneously measure $m_A$, $m_B$, and $m_C$, the $M_{T2}$ subsystems method relies on the measurement of three independent endpoints: $E_{210}$ ($M_{T2}^\text{max}(\tilde{m}_C=0)$ of the $C^1E^1C^2E^2$ system), $E_{221}$ ($M_{T2}^\text{max}(\tilde{m}_C=0)$ of the $D^1E^1D^2E^2$ system), and $E_{im}$ (endpoint of the invariant mass of the visible $D^1E^1$ or $D^2E^2$ systems).  These endpoints can be expressed as functions of the physical masses in the decay chain. Following the naming convention in Ref.~\cite{Burns:2008va},
\begin{eqnarray}
\label{E221}
E_{221} &=& \frac{m_A^2 - m_B^2}{m_A}, \\ 
E_{210} &=& \sqrt{ \frac{\left(m_A^2 - m_C^2\right)\left(m_B^2-m_C^2\right)}{m_A^2} }, \label{E210}\\
E_{im} &=& \sqrt{ \frac{\left(m_A^2 - m_B^2\right)\left(m_B^2-m_C^2\right)}{m_B^2} } \label{Eim}.
\end{eqnarray}
Given a measurement of the endpoints $E_i$, where $i=221$, 210, or $im$, each with an associated uncertainty, $\sigma_{E_i}$, one can, in principle, invert Eqs.~(\ref{E221})-(\ref{Eim}) to solve for $m_A$, $m_B$, and $m_C$ as a function of the three $E_i$'s.  However, because the measured endpoints may not have exactly the expected values, the system of equations may not be invertible without introducing large uncertainties.  For this reason, we study the $M_{T2}$ subsystems method, as we did the $M_{T2}$ kink method, in order to estimate how massive must the dark matter be in order to be distinguished from anomalous neutrino production.  

Instead of inverting Eqs.~(\ref{E221})-(\ref{Eim}), we choose to perform a $\chi^2$ fit using the measured values of $E_i$, $\sigma_{E_i}$, and the expected values of the endpoints as a function of the physical masses as expressed in Eqs.~(\ref{E221})-(\ref{Eim}).  This $\chi^2$ function can be minimized with respect to  $m_A$, $m_B^2$, and $m_C^2$, yielding the best estimates $\hat{m}_A$, $\hat{m}_B^{2}$, and $\hat{m}_C^{2}$ for every set of $E_i$ and $\sigma_{E_i}$.  In the fit, we allow $m_B^2$ and $m_C^2$ to float negative but constrain $m_A$ to be positive.  We assume that the uncertainties associated with the three $E_i$'s are uncorrelated.  

To estimate the uncertainties associated with our ability to measure $m_C$, we generate fifty thousand pseudo-experiments, each with a set of three endpoint measurements, $E_i$, and each endpoint with the same uncertainty, $\sigma_{E_i}$, for simplicity.  A $\chi^2$ function is minimized for each pseudo-experiment, negative values of $\hat{m}_C^{2}$ are set to zero (since they are unphysical), and the square root of the positive values of $\hat{m}_C$ are histogramed.  We find that the center of this distribution is centered about the physical mass $m_C$ and integrate about this center to find the 95\% C.L. associated with the uncertainty of $m_C$.

\begin{figure}[htbp]
\begin{center}
\subfigure[]{\label{sig14}\includegraphics[width=0.45\textwidth]{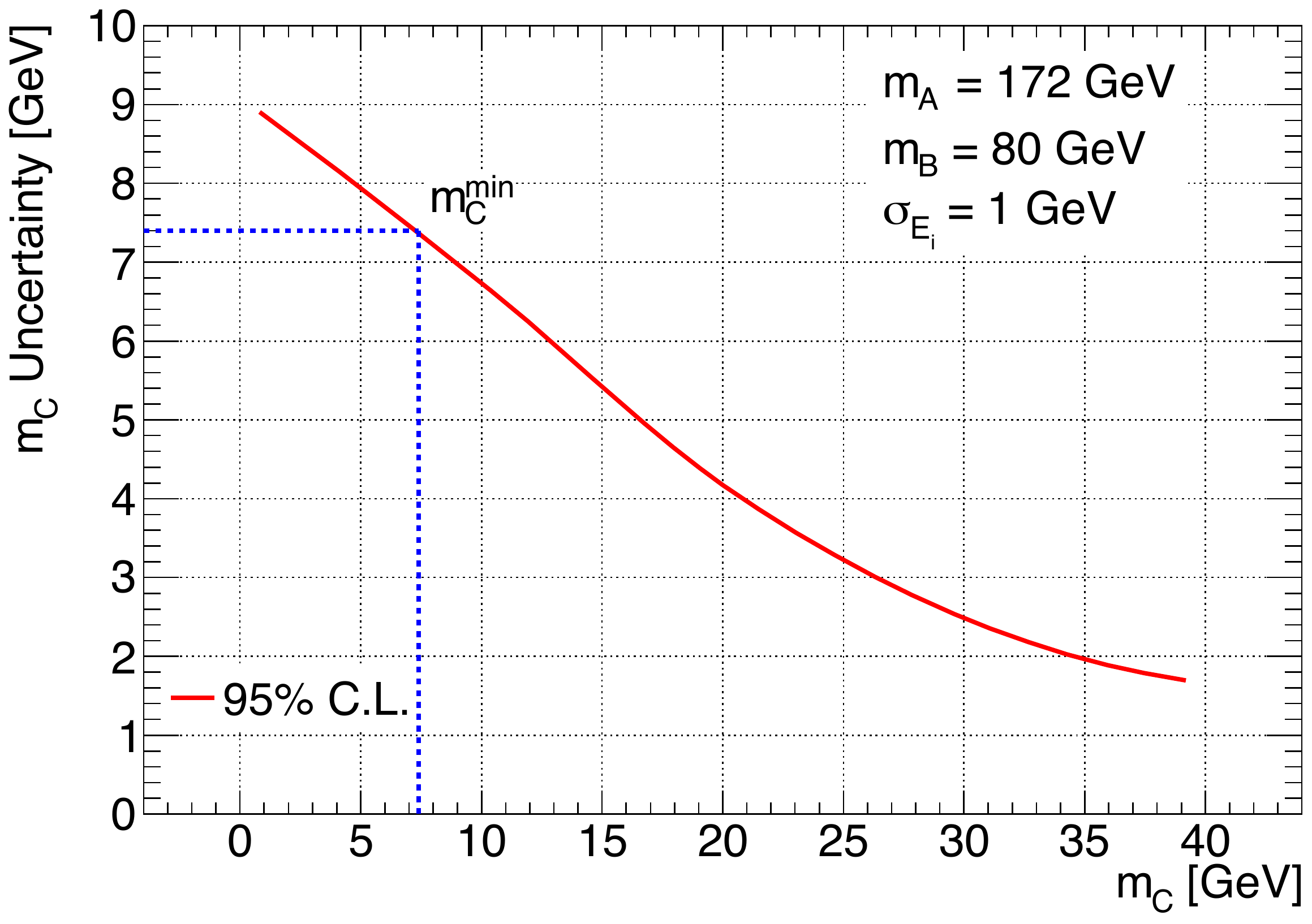}}
\subfigure[]{\label{sig54}\includegraphics[width=0.45\textwidth]{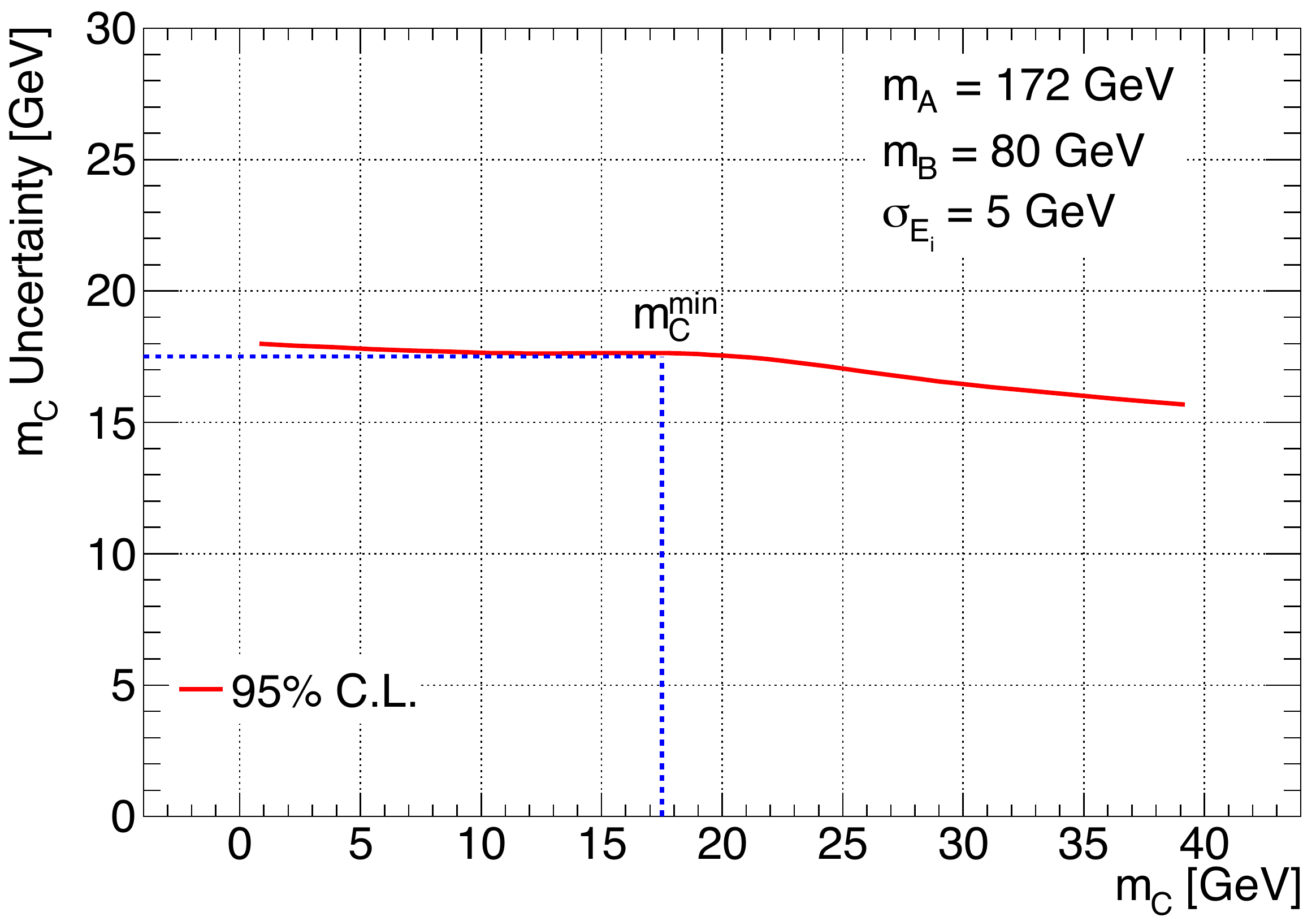}}
\caption{The 95\% C.L. for the mass of $m_C$ for $m_A=172$ GeV and $m_B=80.4$ GeV with (a) $\sigma_{E_i}=1$ GeV and (b) $\sigma_{E_i}=5$ GeV, using the $M_{T2}$ subsystems method.  The variable $m_C^\text{min}$ is the value of mass of $m_C$ at which it can be distinguished from zero at 95\% C.L.}
\label{subsystems1}
\end{center}
\end{figure}

\begin{figure}[htbp]
\begin{center}
\subfigure[]{\label{sig15}\includegraphics[width=0.45\textwidth]{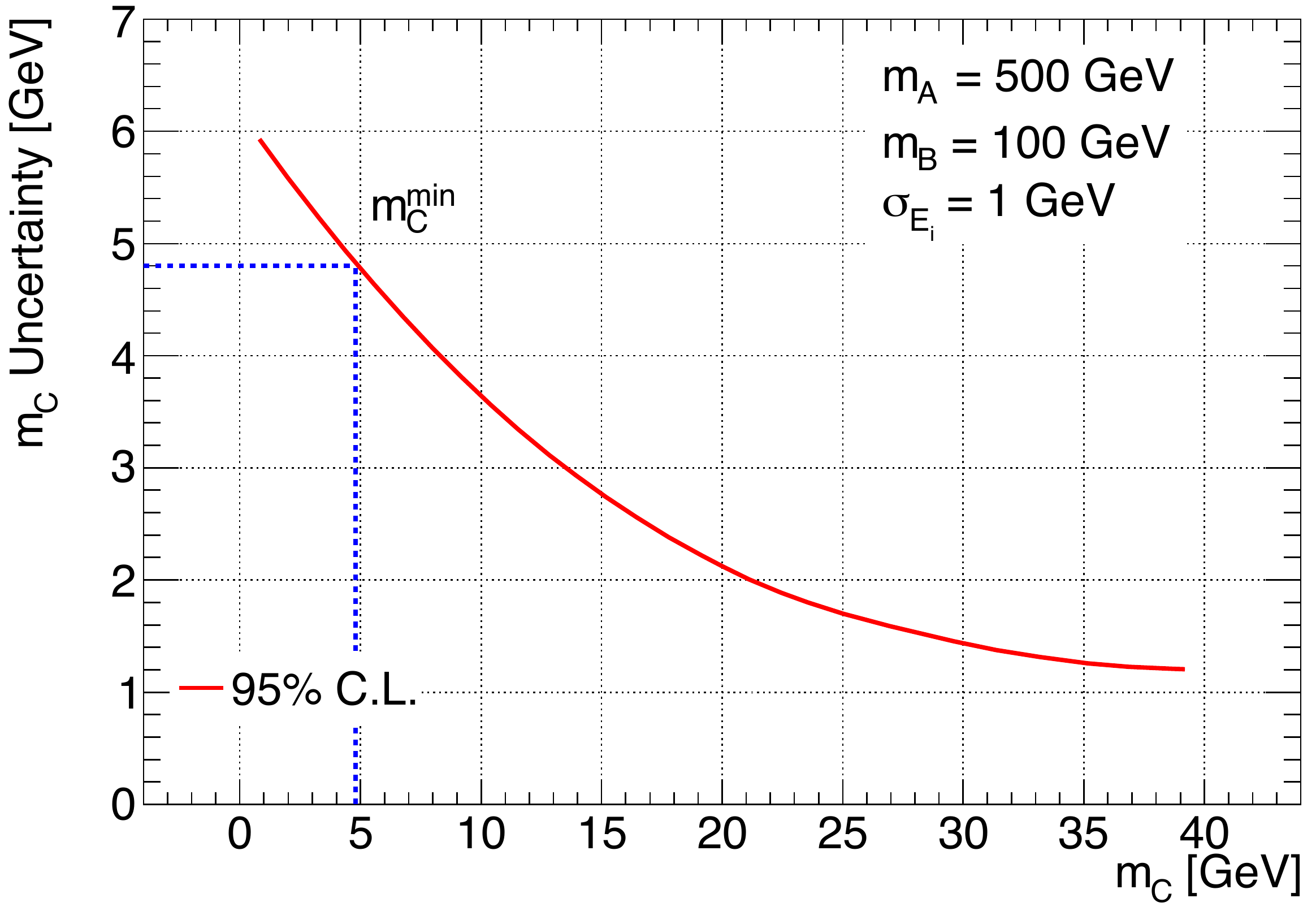}}
\subfigure[]{\label{sig55}\includegraphics[width=0.45\textwidth]{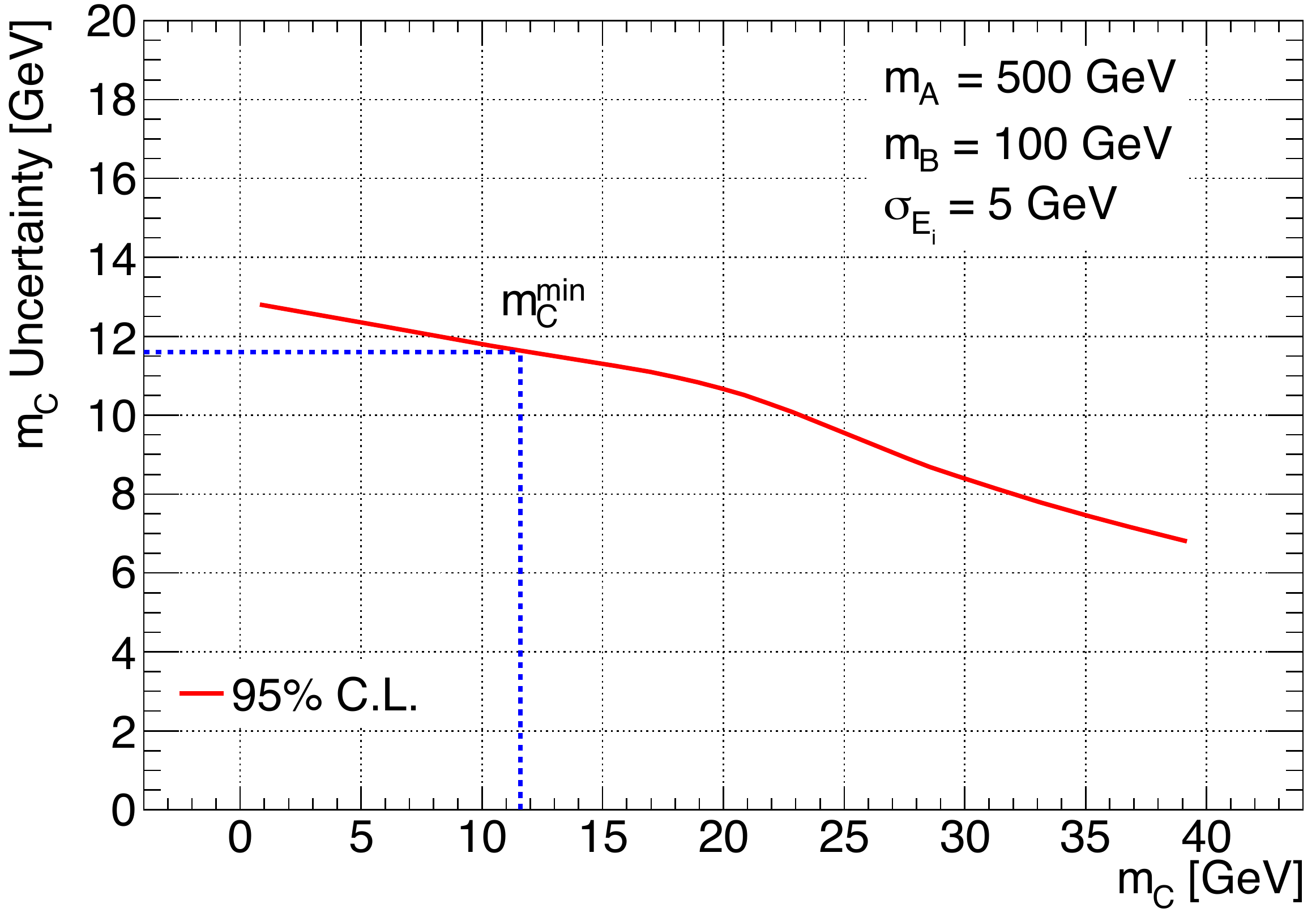}}
\caption{As in Fig.~\ref{subsystems1}, for $m_A=500$ GeV $m_B=100$ GeV.}
\label{subsystems2}
\end{center}
\end{figure}

\begin{figure}[htbp]
\begin{center}
\subfigure[]{\label{sig16}\includegraphics[width=0.45\textwidth]{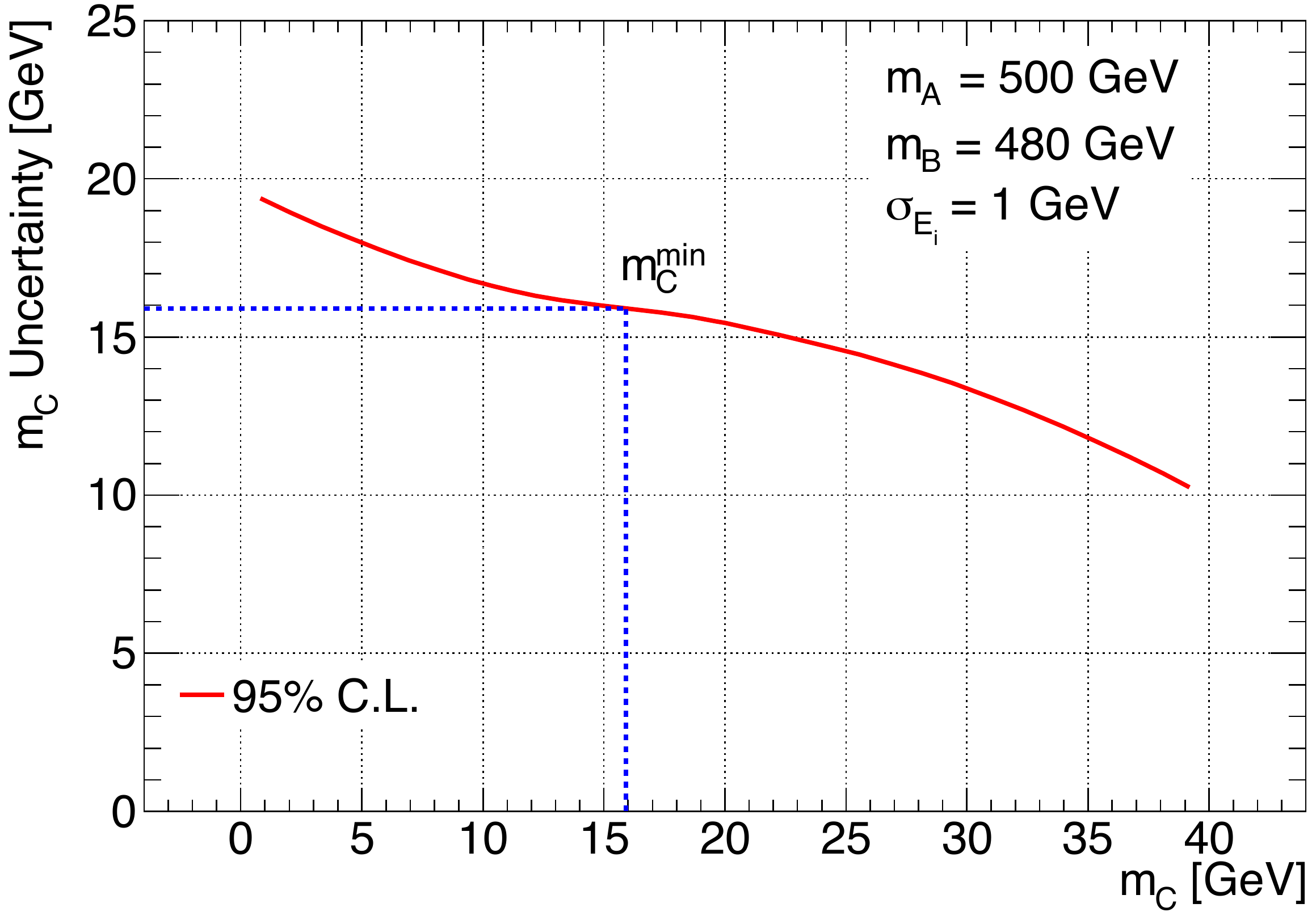}}
\subfigure[]{\label{sig56}\includegraphics[width=0.45\textwidth]{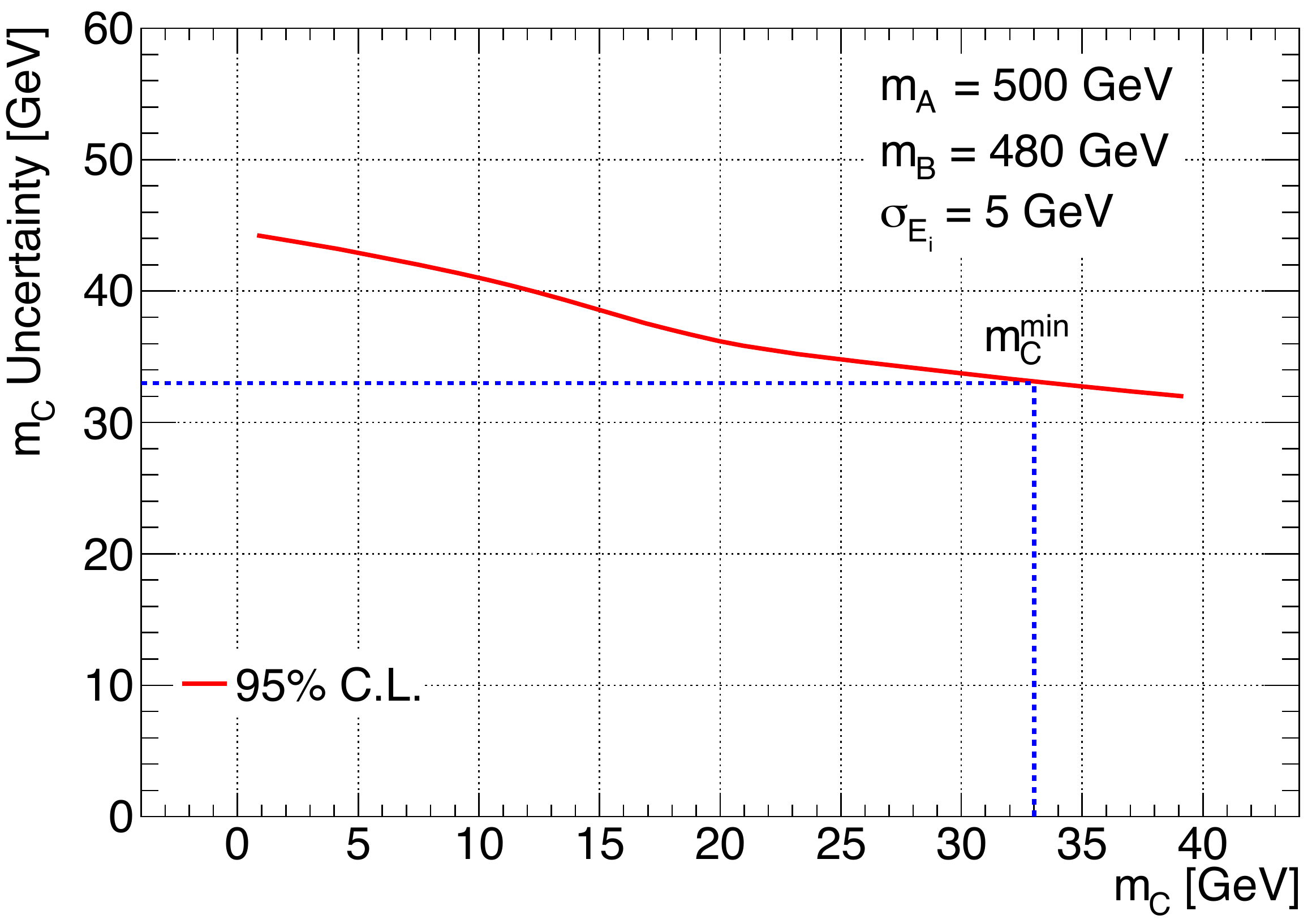}}
\caption{As in Fig.~\ref{subsystems1}, for $m_A=500$ GeV $m_B=480$ GeV.}
\label{subsystems3}
\end{center}
\end{figure}

In order to compare these results with those found using the $M_{T2}$ kink method, we produce similar plots as in Figs.~\ref{kink1}-\ref{kink3}, using the $M_{T2}$ subsystems method.  These are depicted in Figs.~\ref{subsystems1}-\ref{subsystems3}.  We find very similar results for the value of $m_C^\text{min}$ between the $M_{T2}$ subsystems and kink methods. This makes us confident that the procedure described in Appendix~\ref{apa} is sufficient to simulate the correlations between different $M_{T2}$ endpoints as a function of $\tilde{m}_C$.  

\begin{figure}[htbp]
\begin{center}
\subfigure[]{\label{sig1}\includegraphics[width=0.45\textwidth]{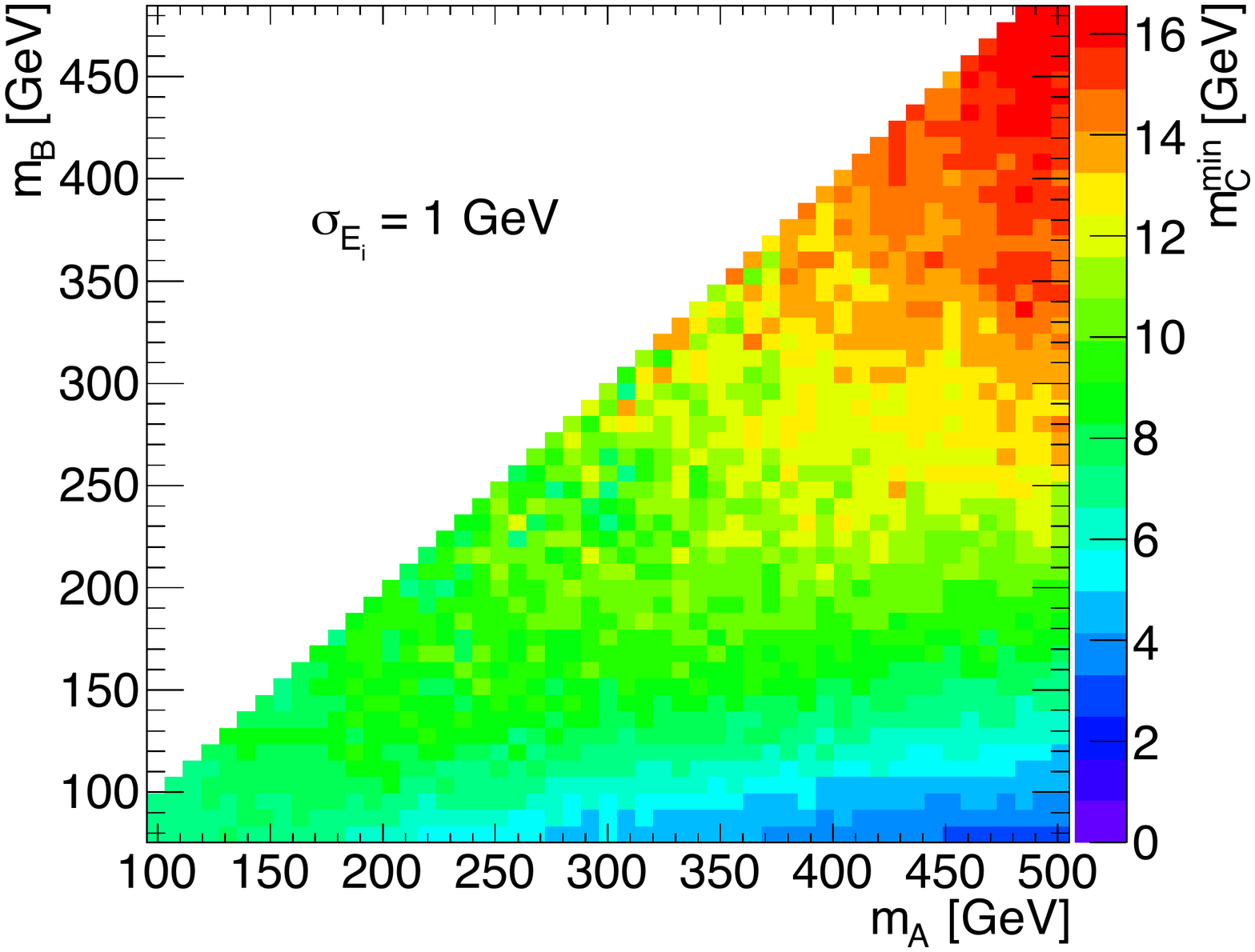}}
\subfigure[]{\label{sig5}\includegraphics[width=0.45\textwidth]{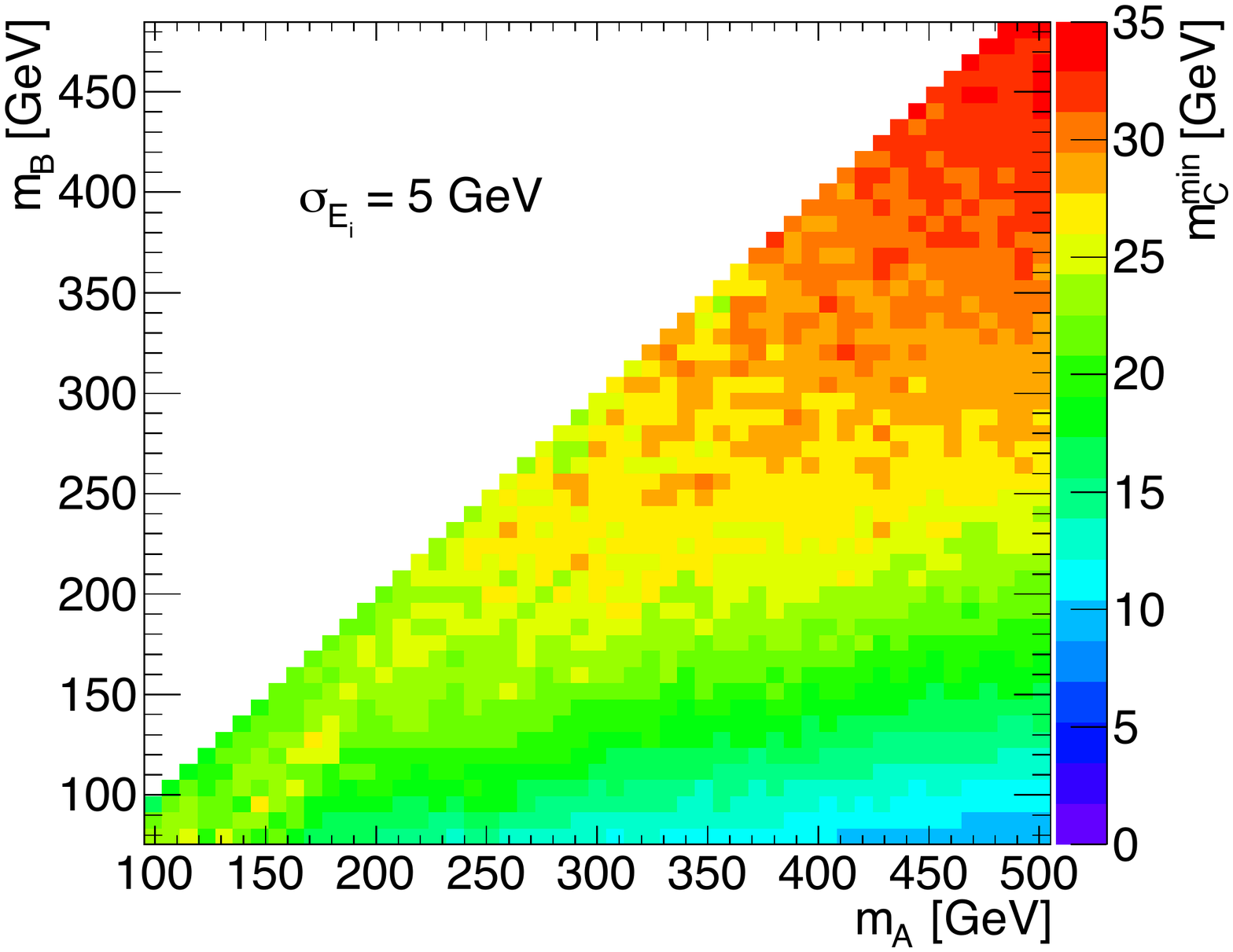}}
\caption{The values of $m_C^\text{min}$, as a function of $m_A$ and $m_B$, when (a) $\sigma_{E_i}=1$ GeV and (b) $\sigma_{E_i}=5$ GeV using the $M_{T2}$ subsystems method.  }
\label{subsystems2D1}
\end{center}
\end{figure}

We estimate $m_C^\text{min}$ as a function of both $m_A$ and $m_B$, as shown in Fig.~\ref{subsystems2D1}, using the $M_{T2}$ subsystems method.  We find when the masses of the parent and intermediate particles are close, the value of $m_C^\text{min}$ is higher as compared to when $m_A$ and $m_B$ have a large mass difference.  We choose not to compute the two dimensional plots of $m_C^\text{min}$ using the $M_{T2}$ kink method, as in Fig.~\ref{subsystems2D1}, since creating the pseudo-data to make the $M_{T2}^\text{max}(\tilde{m}_C)$ distributions is  computationally expensive, and would require a sizable amount of running time.   Again, these results for the value of $m_{C}^\text{min}$, given the values of $m_A$, $m_B$, and $\sigma_E$, can be understood as a lower bound on a measured non-zero mass of dark matter. 

To study how the value of $m_C^\text{min}$ changes as one increases $\sigma_E$ past 5 GeV, we repeat the $M_{T2}$ subsystems analysis for values of $\sigma_E=10$, 15, and 20 GeV.  We find that the value of $m_C^\text{min}$ increases roughly quadratically with an increase in $\sigma_E$, and we observe no effects of saturation.  For example, when $m_A = 172$ GeV, $m_B = 80$ GeV, and $\sigma_E$ = 1, 5, 10, and 15 GeV, the values of $m_C^\text{min}$ are roughly 8, 17, 70, 150 GeV, respectively.

% CONCLUSION
\section{Conclusion}
\label{conclusion}

The observation of new physics events with large missing transverse energy at the LHC would potentially serve as evidence for the production of dark matter. A crucial step toward verifying such evidence is the measurement of the masses of the would-be dark matter particles, i.e., the invisible particles. If, say, an excess of invisible particles is discovered and their masses are found to be consistent with zero, one can either conclude that there are new light massive invisible particles, their masses obfuscated by experimental resolution, or that there is new physics in the neutrino sector~\cite{Chang:2009dh}.  As hinted by the CoGeNT~\cite{Aalseth:2010vx}, DAMA/LIBRA~\cite{Bernabei:2008yi,Bernabei:2010mq}, and CRESST~\cite{Angloher:2011uu} experiments, dark matter may have a mass of $\mathcal{O}(\text{10 GeV})$. Coincidently, assuming that dark matter is produced at hadron colliders in a $t\bar{t}$-like event, i.e., the ``best-case" decay topology and considering optimistic experimental uncertainties, we find that dark matter must have a mass greater than $\mathcal{O}(\text{10 GeV})$ such that it can be distinguished from neutrinos at 95\% C.L. using $M_{T2}$-based methods. In general, the uncertainty associated with measuring the mass the invisible particles increases (decreases) as the mass decreases (increases).  Our results suggest that, at the LHC, it may prove very challenging to distinguish light dark matter from neutrinos through mass measurements alone if the dark matter weighs around 10~GeV or less.

As seen in Fig.~\ref{subsystems2D1}, the uncertainty associated with the measurement of the invisible particles' mass increases as the mass $m_B$ of the intermediate particles  is close to the mass $m_A$ of the parent particles, and one finds the inverse effect if the intermediate particles are much lighter than the parent particles.  In other words, a more precise measurement of the mass of the invisible particles can be made when the intermediate particles have a larger momentum in the center-of-mass frame of the two parent particles.  Furthermore, the uncertainty associated with the measurement of the invisible particles' mass also increases as the parent particles' mass increases, for fixed $m_B/m_A$.  These features are due to the fact that the kink structure of the $M_{T2}^\text{max}(\tilde{m}_C)$ distribution becomes more pronounced when there is a large mass difference between the parent and intermediate particles.  A similar scenario occurs when the intermediate particle is off-shell, such that $m_B > m_A > m_C$, i.e., the kink structure becomes more pronounced~\cite{Burns:2008va}.  While a more pronounced kink structure will increase the precision at which one can measure the masses of particles involved in a given decay chain, it is not generally expected that it will change the shape of the uncertainty associated with $m_C$ as a function of the physical masses.  

We observe a rough quadratic relationship between between the values of $\sigma_E$ and $m_C^\text{min}$.  Using the $M_{T2}$ subsystems method, we let $m_A=172$ GeV, $m_B=80$ GeV, and $\sigma_E$ = 1, 5, 10, and 15 GeV, and find the associated values of $m_C^\text{min}$ to be roughly 8, 17, 70 and 150 GeV, respectively.   Beyond $\sigma_E = 15$ GeV, the uncertainty on the mass of the invisible particle becomes very large. 

\begin{comment}
While the shapes of the distributions found in Fig.~\ref{subsystems2D1} are different, if one assumes that Fig.~\ref{sig1} and Fig.~\ref{sig5} are related by a multiplicative constant, one can make a rough estimation for what happens if $\sigma_{E_i}$ is increased beyond 5 GeV.  If one sets $\sigma_{E_i} = \mu\times 1$ GeV, then we find roughly that the result for $m_C^\text{min}$ will be the product of $\mu$ and the values of $m_C^\text{min}$ found in Fig.~\ref{sig1}.  We find this rough linear relation even holds true up to value of $\sigma_{E_i}=20$ GeV, where the values of $m_C^\text{min}$ are about 20 times the values found in Fig.~\ref{sig1}.  
\end{comment}

The individual events that populate the region about an $M_{T2}$ endpoint for a given value of the ansatz mass are, to a good approximation, the same events that populate the endpoint for another nearby value of the ansatz mass.  As shown in Fig.~\ref{mt2endpoints} (see Appendix~\ref{apa}), the values of $M_{T2}^\text{max}$ are not randomly distributed about the theoretical expectation, and while there are systematic uncertainties associated with $M_{T2}$ endpoint-finding procedure, there can be significant correlations between the measured values $M_{T2}^\text{max}(\tilde{m}_C)$ for adjacent values of $\tilde{m}_C$. For this reason, one cannot simply  fit the measured value of $M_{T2}^\text{max}$ using Eq.~(\ref{mt2max}) while treating the individual values of $M_{T2}^\text{max}$ as independent measurements.  In our analysis, we take this correlation into account when fitting with Eq.~(\ref{mt2max}) and marginalize over the other fitting parameters in order to correctly quote the uncertainty on the mass of the invisible particles. The majority of analyses in the literature ignore the correlation between values of $M_{T2}^\text{max}$ for adjacent values of $\tilde{m}_C$, e.g., Refs.~\cite{Cho:2007qv, Cho:2007dh, Choi:2010dk, Belanger:2011ny, Nojiri:2008hy}.  If these correlations are not taken into account when using the $M_{T2}$ kink method, the uncertainties associated with the measured masses can be underestimated.   

Other $M_{T2}$-based methods and topologies could have been considered in our analysis. All methods capable of extracting the mass of the invisible particles, however, take advantage of singularities in the kinematic phase space~\cite{Kim:2009si, Cheng:2008hk}.  Though some attention has been paid to other methods that can, in principle, provide more precise mass measurements~\cite{Matchev:2009fh, Konar:2009wn}, we speculate that the limitations of the $M_{T2}$ kink and subsystems methods, discussed here in some detail, will be similar to the limitations of other methods as far as determining the mass of light invisible particles is concerned.   Because our results were so similar for the $M_{T2}$ kink and subsystems methods, it is possible that these trends are independent from the method used to extract the mass of the invisible particles.

\begin{acknowledgments}
The authors are grateful to Spencer Chang, Nicholas Eggert, KC Kong, and Konstantin Matchev for useful conversations and feedback.  The work of AdG is sponsored in part by the DOE grant \# DE-FG02-91ER40684.  ACK is supported in part by the Department of Energy Office of Science Graduate Fellowship Program (DOE SCGF), made possible in part by  the American Recovery and Reinvestment Act of 2009, administered by ORISE-ORAU under contract no. DE-AC05-06OR23100.
\end{acknowledgments}

% APPENDIX A
\appendix
\section{Correlations between $M_{T2}$ endpoints}
\label{apa}

\begin{figure}[htbp]
\begin{center}
\includegraphics[width=0.5\textwidth]{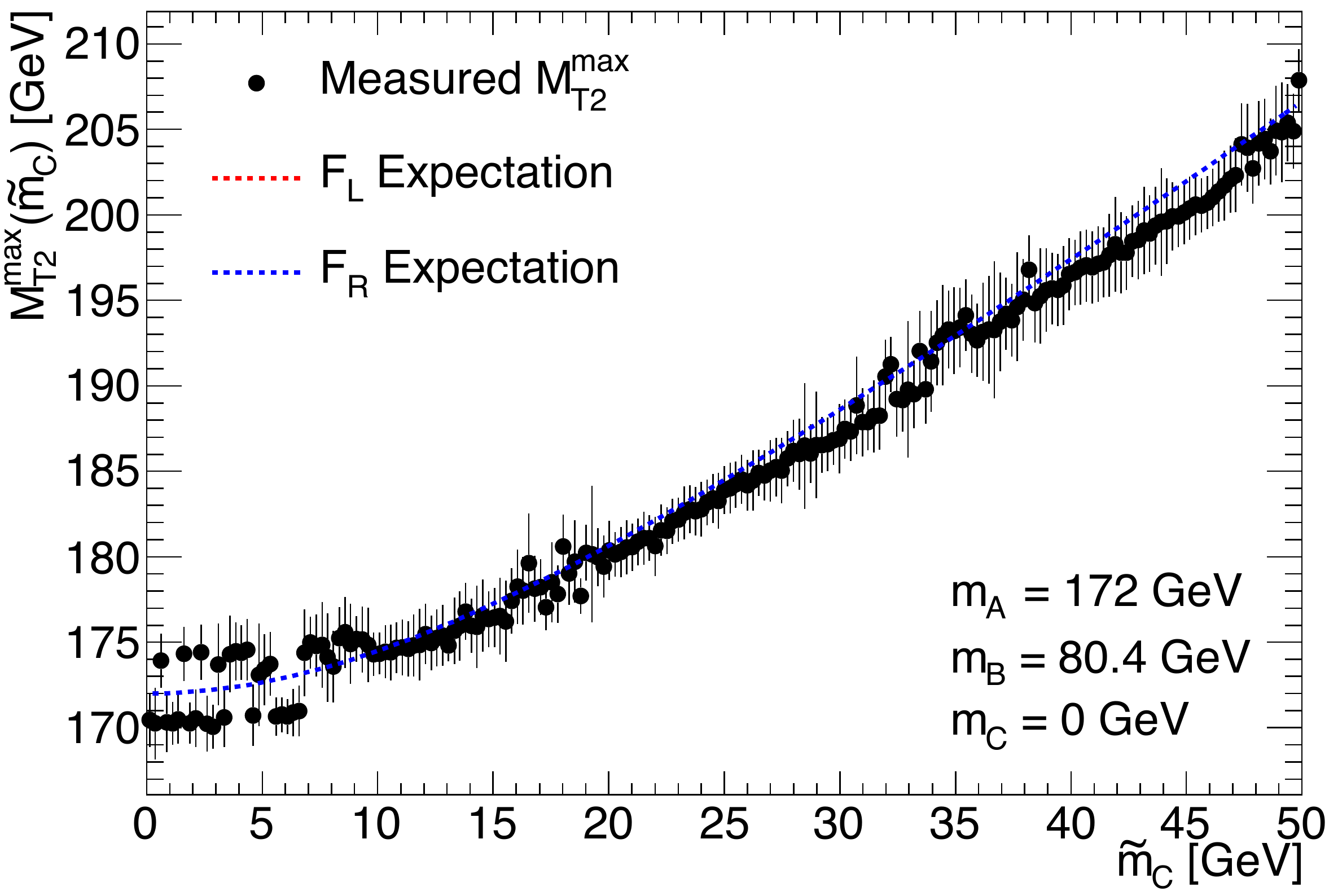}
\caption{The fitted endpoints of $M_{T2}$ at the generator level with one hundred thousand events for $m_A=172$ GeV, $m_B=80.4$ GeV, and $m_C=0$ GeV using an unbinned endpoint fitting procedure.  The dashed blue and dashed red curves are the theoretical expectations (the red dashed line is not shown since $m_C=0$ GeV). Note that the points are not randomly distributed about the theoretical expectation.}
\label{mt2endpoints}
\end{center}
\end{figure}

To understand the correlation between endpoint measurements, we generate a {\sc madgraph5}~\cite{Alwall:2011uj} sample of one hundred thousand SM $pp\rightarrow t\bar{t}$ events at $\sqrt{s} = 7$ TeV, where the mass of the top quark is 172 GeV (corresponding to $m_A$), and the mass of the $W$ is 80.4 GeV (corresponding to $m_B$).  At the generator level, we fit the endpoint of a given $M_{T2}$ distribution with the following four-parameter  probability density function:
\begin{equation}
\rho(x; A, k_L, k_R, x_0) = \left\{ \begin{array}{ll}   Ae^{-k_R(x-x_0)}, & x_\text{min} < x < x_0, \\ A e^{-k_L(x-x_0)}, & x_0 < x < x_\text{max}, \end{array} \right. 
\end{equation}
where $x$ is a dummy variable for the $x$-axis of an $M_{T2}$ distribution, $A$ is a normalization constant, $x_0$ is taken to be the location of $M_{T2}^\text{max}$, and $k_L$ and $k_R$ are the exponential slopes on the left and right sides of the kink, respectively.  We assume no combinatorial background associated with which jet is associated with which decay branch.  The values of $x_\text{min}$ and $x_\text{max}$ are chosen {\it a priori} to be certain values about the expected value of $M_{T2}^\text{max}$.  While picking values of $x_\text{min}$ and $x_\text{max}$ based off a theoretical expectation introduces a bias, we are only interested in extracting information concerning a rough estimation of how correlated adjacent endpoint measurements are as a function of $\tilde{m}_C$.  For each $M_{T2}(\tilde{m}_C)$ distribution, we perform an unbinned log-likelihood fit, where the log-likelihood function is defined using events with a value of $M_{T2}$ between $x_\text{min}$ and $x_\text{max}$,
\begin{equation}
\ln \mathcal{L} = \displaystyle\sum_{{x_\text{min} < x_i < x_\text{max}}} \ln\rho(x_i; A, m_L, m_R, x_0).
\end{equation}
The extremization of $\ln\mathcal{L}$ is performed with {\sc minuit}~\cite{James:1975dr}, and the results of the fits are shown in Fig.~\ref{mt2endpoints}.  The uncertainty associated with $x_0$ is found by marginalizing over the uncertainties for $A$, $m_L$, and $m_R$ and varying the log-likelihood about the minimum function by a value of 0.5 as a function of $x_0$.  

While this particular {\sc madgraph5} sample vastly oversimplifies the type of distributions one would have to work with at a collider experiment, we only wish to extract information concerning the correlation between adjacent endpoints.  To estimate the correlation between $M_{T2}$ endpoints and to create pseudo-data, we need to simulate a distribution that properly mimics the one in Fig.~\ref{mt2endpoints} while avoiding having to make thousands of {\sc madgraph5} samples.  To create pseudo-data that contain the correlations we see in Fig.~\ref{mt2endpoints}, we need to quantify how much each endpoint measurement is correlated with others.  

\begin{figure}[htbp]
\begin{center}
\subfigure[]{\label{NNC00}\includegraphics[width=0.45\textwidth]{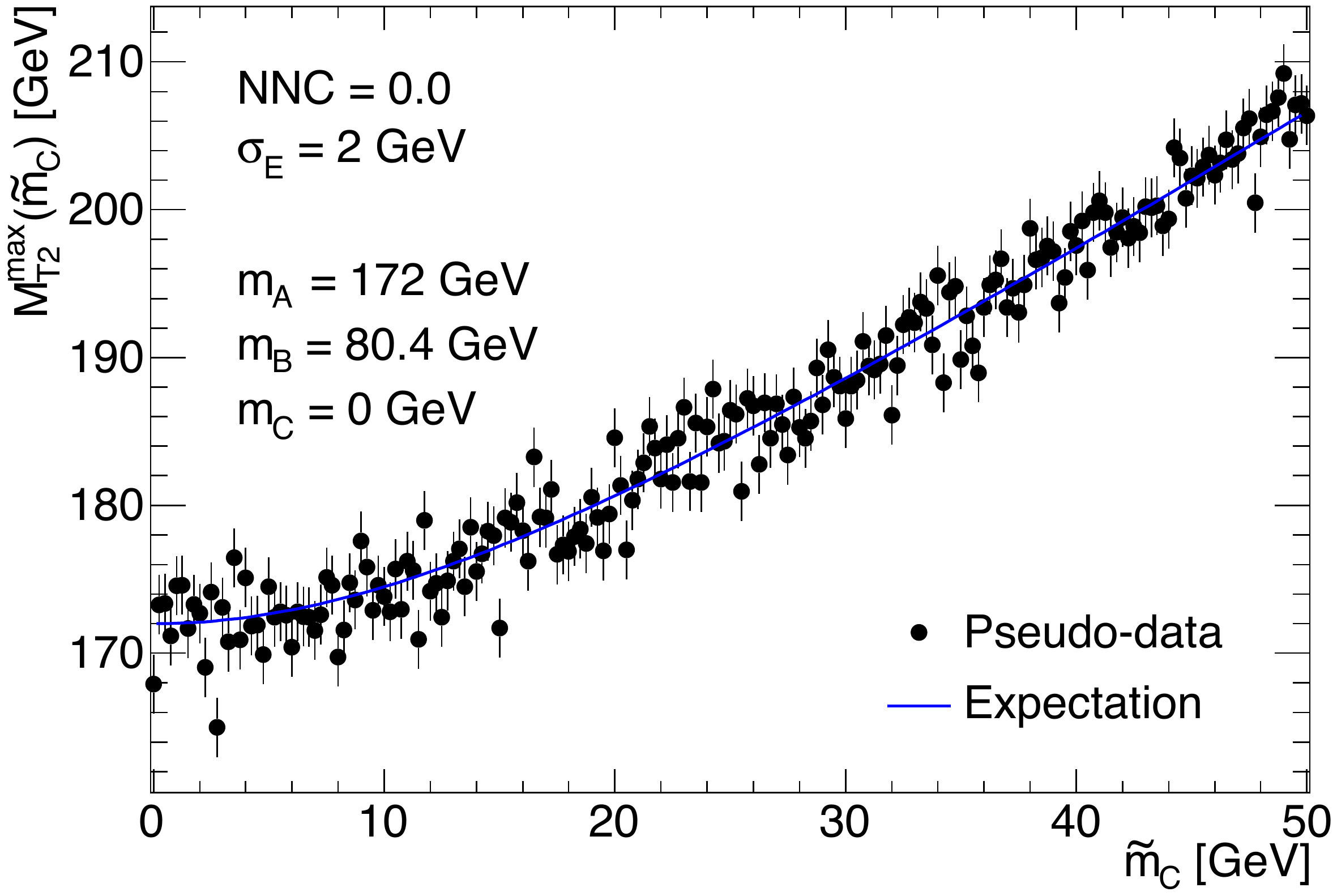}}
\subfigure[]{\label{NNC04}\includegraphics[width=0.45\textwidth]{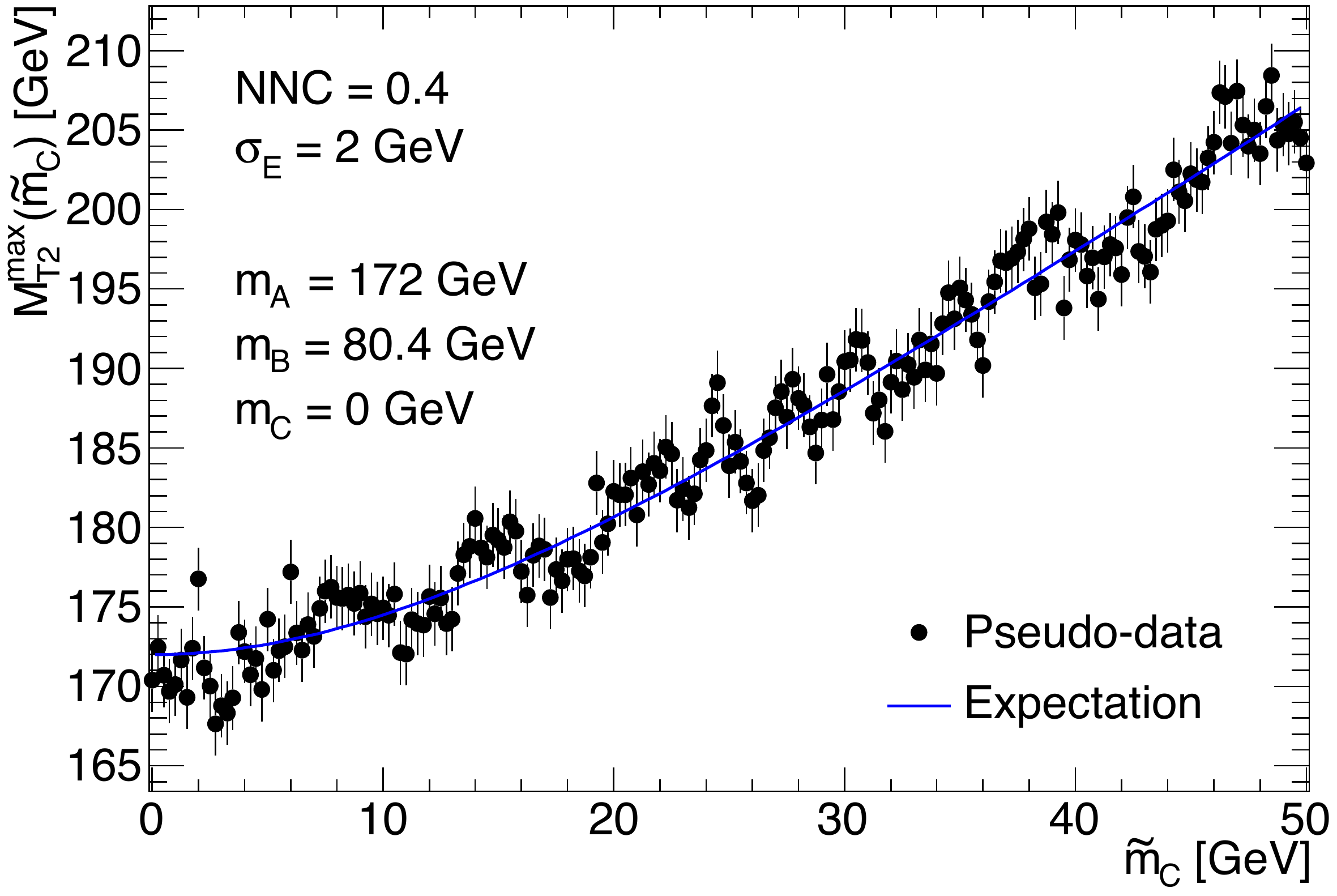}}
\subfigure[]{\label{NNC05}\includegraphics[width=0.45\textwidth]{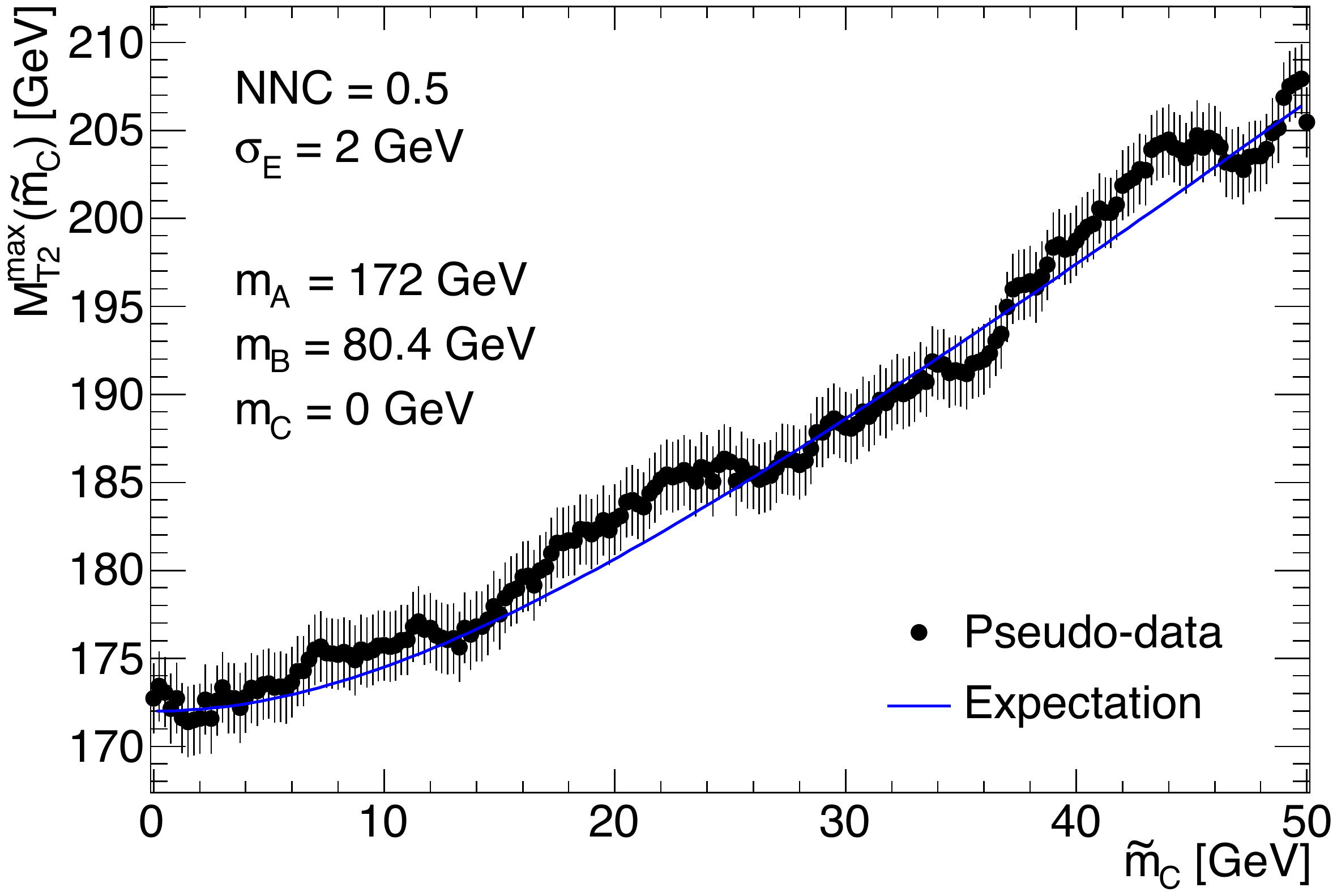}}
\caption{Sample simplified pseudo-data for $M_{T2}^\text{max}$ distribution with for $m_A = 172$ GeV, $m_B = 80.4$ GeV, $m_C = 0$ GeV, $\sigma_E = 2$ GeV, and (a) NNC = 0, (b) NNC = 0.4, and (c) NNC = 0.5. } 
\label{NNC}
\end{center}
\end{figure}

We employ a simplified model of nearest-neighbor correlations.  In this method, a single endpoint has its own uncertainty, $\sigma_i$, and is positively correlated with the endpoint immediately to the left (except for the left-most endpoint, which is taken to have no correlations with any other endpoint).  For simplicity, we consider that all measurements of $M_{T2}^\text{max}$ have the same uncertainty, $\sigma_E$.  We create a covariance matrix, which only includes nearest-neighbor correlations:  
\begin{equation}
V_{ij} = \left\{ \begin{array}{ll} \sigma_E^2, & i=j \\ \text{NNC} \times \sigma_E^2, & |i-j|=1 \\ 0, & \text{otherwise},  \end{array} \right.
\end{equation}
Here, NNC is the nearest-neighbor correlation factor.  With this error matrix, one can iteratively generate pseudo-data that resemble a type of distribution we find with a {\sc madgraph5} sample and a simple fitting procedure.  We begin with the first point, $n_0$, at $\tilde{m}_C=0$ GeV, allowing it to be Gaussian-distributed about the theoretical expectation, $\mu_0$, with an error of $\sigma_E$.  The next point, $n_1$, is Gaussian-distributed about the expectation $\mu_1$ with uncertainty $\sigma_E$ but also is positively correlated to the the previous point.  These endpoints are random variables sampled from the following probability density function (up to an overall normalization factor):
\begin{equation}
\rho = \left\{ \begin{array}{ll} e^{-(n_i-\mu_i)^2/\sigma_E^2}, & i=0 \\  e^{-(n_i-\mu_i)^2/\sigma_E^2} \times  e^{-(n_i-\mu_i)(n_{i-1} - \mu_{i-1})V_{i,i-1}^{-1}}, & i>0,  \end{array} \right.
\end{equation}
Examples for the simplified pseudo-data for $M_{T2}^\text{max}$ are shown in Fig.~\ref{NNC}. 

\begin{figure}[htbp]
\begin{center}
\includegraphics[width=0.5\textwidth]{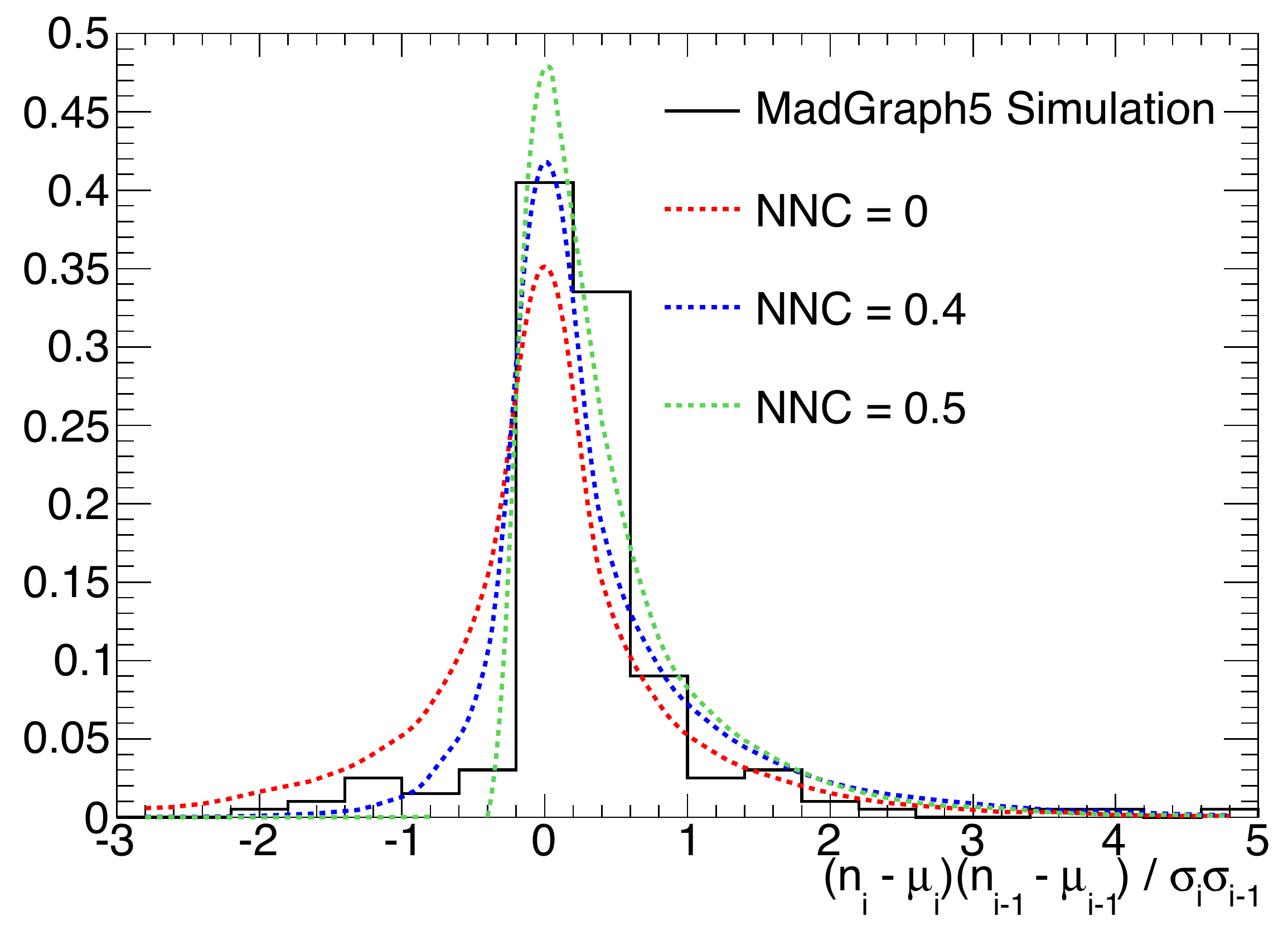}
\caption{The variable $R$ as defined in Eq.~(\ref{Rdef}) for the {\sc madgraph5} distribution as in Fig.~\ref{mt2endpoints} (black), and three values of the nearest-neighbor correlation factor, NNC.  All distributions are normalized to unity. }
\label{mt2corr}
\end{center}
\end{figure}

While comparing the {\sc madgraph5} and simplified pseudo-data plots by eye is suitable for our purposes, we wish to compare them more quantitatively.  We can create a test statistic that estimates how much a given endpoint, $n_i$, correlates with the endpoint immediately to its left, $n_{i-1}$:
\begin{equation}
\label{Rdef}
R = \frac{(n_i - \mu_i)(n_{i-1}-\mu_{i-1})}{\sigma_i \sigma_{i-1}},
\end{equation}
where $\mu_i$ and $\mu_{i-1}$ are the theoretical expectations (given the physical masses of the system) for the $M_{T2}^\text{max}$ distribution for bin $i$ and the bin immediately to the left, $i-1$.  If each endpoint is statistically uncorrelated with the one immediately to the left, the distribution for $R$ should be symmetric about zero.  Fig.~\ref{mt2corr} shows a histogram of the values of $R$ for the {\sc madgraph5} sample and the probability density functions for pseudo-data for different values of NNC (these lines are determined by averaging over many generated pseudo-data).  Guided by these distributions, we choose a value of NNC = 0.5 to perform the study discuss in Sec.~\ref{kinkmethod}.  We choose this value because we interpret that the random jumps in Fig.~\ref{mt2endpoints} for $0<\tilde{m}_C<10$ GeV are innocuous qualities of our endpoint fitting procedure, and subsequently the histogramed values of $-2 < R < 0$ in Fig.~\ref{mt2corr} for the {\sc madgraph5} sample are not sampled from the true probability density function for $R$.  Values larger than NNC = 0.5 where not studied because they lead to pseudo-data that deviated too much from the theoretical expectation, which is a feature not found in Fig.~\ref{mt2endpoints}.  Finally, we find that the correlation between adjacent values of $\tilde{m}_C$ is not particularly sensitive to the physical masses of the particles in the decay chain, including $m_C$.   

\bibliography{bib}{}

\end{document}